# INTRODUCTION*

M. Shifman
*William I. Fine Theoretical Physics Institute*
*University of Minnesota, Minneapolis, MN 55455 USA*
*shifman@umn.edu*

*When destination becomes destiny...*

— 1 —

Five decades — from the 1920s till 1970s — were the golden age of physics. Never before have developments in physics played such an important role in the history of civilization, and they probably never will again. This was an exhilarating time for physicists.

The same five decades also witnessed terrible atrocities, cruelty and degradation of humanity on an unprecedented scale. The rise of dictatorships (e.g., in Europe, the German national socialism and the communist Soviet Union) brought misery to millions. *El sueño de la razón produce monstruos...*

In 2012, when I was working on the book *Under the Spell of Landau*, [1] I thought this would be my last book on the history of theoretical physics and the fate of physicists under totalitarian regimes (in the USSR in an extreme form as mass terror in the 1930s and 40s, and in a milder but still onerous and humiliating form in the Brezhnev era). I thought that modern Russia was finally rid of its dictatorial past and on the way to civility. Unfortunately, my hopes remain fragile: recent events in this part of the world show that the past holds its grip. We are currently witnessing recurrent (and even dangerously growing) symptoms of authoritarian rule: with political opponents of the supreme leader forced in exile or intimidated, with virtually no deterrence from legislators or independent media, the nation's future depends on decisions made singlehandedly. Ob-

* Abridged Introduction to "Physics in a Mad World" (World Scientific, 2015).

serving current events in Russia, I better understand how Nazi Germany or the Stalin-Brezhnev Soviet Union could have happened. The future of Russia at large, and of the Russian intelligentsia in particular, is rather unpredictable at the moment. Alas... it seems that lessons from the past are never obsolete.

Recently I came across a number of remarkable essays written in Russian about the lives of two theoretical physicists in the USSR. Although they did not know each other, they both spent some time in Kharkov (now in Ukraine), and there are many other commonalities in their destinies. These essays can be read as highly instructive detective stories, and I decided that it would be important to familiarize Western readers with them. Thus I returned to the task I had set myself previously and which I had left when I finished *Under the Spell of Landau*.

This collection will tell the captivating stories of the misadventures of two renowned physicists. The first part of the book is devoted to Friedrich (Fritz) Houtermans, an outstanding Dutch-Austrian-German physicist who was the first to suggest that the source of stars' energy is thermonuclear fusion, and who made a number of other important contributions to astrochemistry and geochemistry. In 1935 Houtermans, who was a German communist, fled to the Soviet Union in an attempt to save his life from Hitler's Gestapo. Houtermans took an appointment at the Ukrainian Physico-Technical Institute (also known as Kharkov Fiztech) and worked there for three years with the Russian physicist Valentin P. Fomin. In the Great Purge of 1937, Houtermans was arrested by the NKVD (the Soviet Secret Police, the KGB's predecessor[1]) in December 1937. He was tortured and confessed to being a Trotskyist plotter and Ger-man spy, out of fear from threats against his wife Charlotte.[2] However, by that time Charlotte had already escaped from the Soviet Union to Denmark, after which she went to England and finally the USA. After the Hitler-Stalin Pact of 1939, Houtermans was turned over to the Gestapo in May 1940 and imprisoned in Berlin.

[1]In 1917, shortly after the Bolshevik coup d'etat, the Council of People's Commissars created a secret political police, the Cheka, led by Felix Dzerzhinsky. The Cheka was reorganized in 1922 as the Main Political Directorate, or GPU. In 1934 GPU in its turn was reorganized and became the NKVD. In 1946, all Soviet Commissariats were renamed "ministries." Accordingly, the NKVD was renamed as the Ministry of State Security (MGB) which in 1954 became the USSR Committee for State Security (KGB).

The second part consists of two essays narrating the life story of Yuri Golfand, one of the co-discoverers of supersymmetry, a revolutionary concept in theoretical physics in the twentieth century. In 1973, just two years after the publication of his seminal paper, he was fired from the Lebedev Physics Institute in Moscow. Because of his Jewish origins he could find no job. Under these circumstances, he applied for an exit visa to Israel, but his application was denied. Yuri Golfand became a *refusenik*,[3] and joined the human rights movement, collaborating in this with other prominent physicists, including Andrei Sakharov and Yuri Orlov, among others. Throughout the 1970s, to earn his living he had to do manual casual work, while facing repeated harassment from the KGB. Only 18 years after applying for his exit visa did he obtain permission to leave the country, emigrating to Israel in 1990, shortly before the demise of the Soviet Union.

---

[2]Née Charlotte Riefenstahl (1899-1993), received her doctorate in physics at the University of Göttingen in 1927. In 1930, she left her teaching position at Vassar College (Poughkeepsie, NY, USA) and returned to Germany. After a physics conference in Odessa, USSR in August 1930, , during a trip in the Caucasus organized for the participants, Riefenstahl and Houtermans married, with Wolfgang Pauli and Rudolf Peierls as witnesses to the ceremony.

[3]A group of people treated as political enemies in the USSR in the 1970s and 80s. The only "crime" committed by these people was that they had applied for and been denied exit visas to Israel. Yet, they were treated essentially as criminals: fired from jobs and blacklisted, with no access to work (with the exception of low-paid manual labor), constantly intimidated by the KGB with the threat of arrest or other reprisals. In fact, the most active of them, those who tried to organize and defend their rights, were imprisoned.

— 2 —

Part I of the book presents the English translation of the Victor Yakovlevich Frenkel's monograph *Professor Friedrich Houtermans: Works, Life, Fate*. It was published in 1997 by the St. Petersburg Institute of Nuclear Physics and went largely unnoticed by the general public.[4] Professor Victor Frenkel (1930-1997), a son of the famous Russian physicist Yakov Ilyich Frenkel,[5] was an acclaimed historian of Soviet physics, the author of two academic treatises: one devoted to his father, Yakov Frenkel, and another on Matvei Petrovich Bronshtein (1906-1938), a Soviet theoretical physicist who did pioneering work in quantum gravity and cosmology, and who tragically perished during the Great Purge. In working on the Houtermans book Victor Frenkel extensively used archival materials from many countries, including secret documents which were released in the early 1990s (after the collapse of the Soviet Union) by the FSB, the KGB's successor as modern-day Russia's state security service. This archive was accessible for a while.

In the early 1990s Victor Frenkel made a trip to the USA, to interview Charlotte Houtermans, Friedrich Houtermans' widow, who at that time was still alive and resided in the small town of Northfield, Minnesota. Charlotte provided him with extensive excerpts from her unpublished diaries covering 1937-1938, the last months of the Houtermans' ordeal in Kharkov, and Charlotte's escape from the USSR. The atmosphere of everyday fear permeates each page of these memoirs, which strikingly convey the tone of life in Kharkov

[4]An excellent book about Houtermans' misadventures, which was in the making for about 20 years and complements Frenkel's work, was recently published by Springer Verlag [2].

[5]Yakov Ilyich Frenkel (1894-1952) is known for his outstanding contributions in condensed matter physics. He was the first to propose the notion of "holes," to be interpreted as positively charged quasiparticles. In semiconductor and insulator physics he proposed a theory (1938) which is now referred to as the Poole-Frenkel effect. In the theory of plastic deformations he laid the foundation of what is currently known as the Frenkel-Kontorova-Tomlinson model.

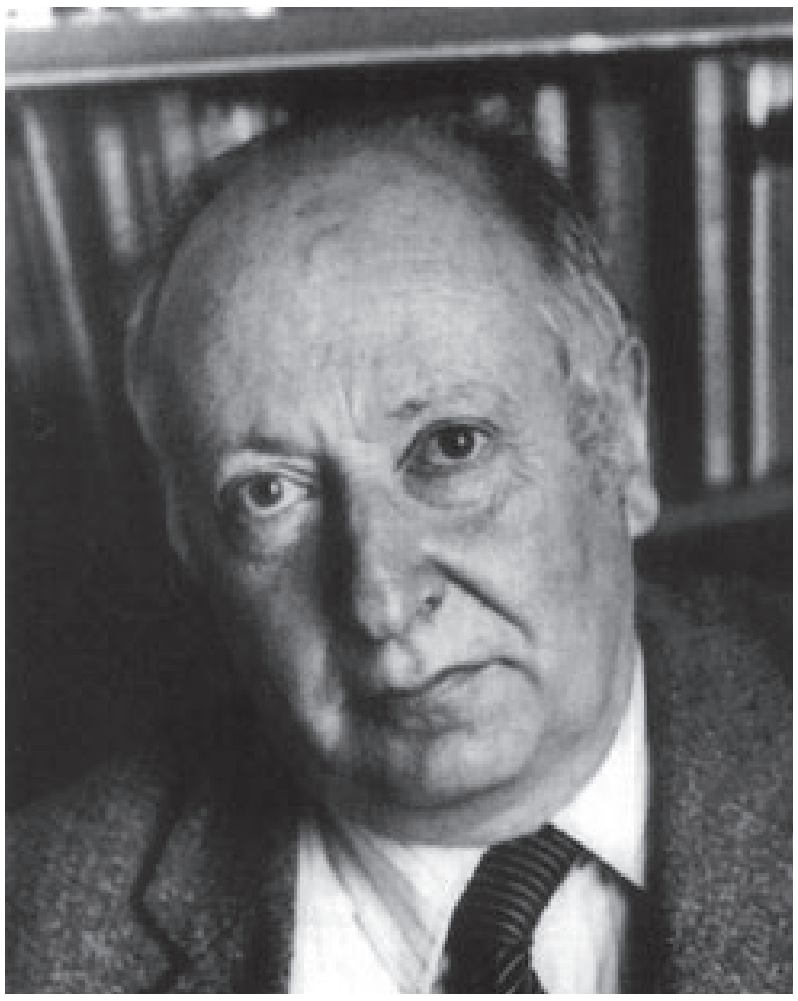
Victor Yakovlevich Frenkel

in the time of Great Purge. Later I will say a few words about the dramatic events at the Ukrainian Physico-Technical Institute: the devastation of its theory group and experimental laboratories, and arrests, arrests, arrests...

Victor Frenkel died with work on the manuscript almost completed but a few gaps still needing to be filled. The completion of the book was carried out by his assistant, Dr. Boris Diakov, whose generous advice on certain issues I was happy to use. He supplemented Victor Frenkel's manuscript with excerpts from talks delivered by Frenkel in Denmark, the USA, Germany and Russia. He also compiled a chronology of Friedrich Houtermans' life.

Copyright to Frenkel's book belongs to his widow, Olga Vladimirovna Cherneva. She kindly agreed to give permission for its English translation. Several chapters in this book are long quotations from the diaries of Charlotte Houtermans. In Frenkel's book they are presented in Russian translation. Of course, it was senseless to translate them back into English.

The English originals of the notes which were provided to Victor Frenkel by Charlotte, as well as the rest of his scientific archive, are buried somewhere in the Ioffe Physico-Technical Institute in St. Petersburg. Despite numerous attempts, I failed to obtain copies of any of these documents from there.

Accordingly, in the beginning of 2014 I set out on course to find Charlotte's daughter, Giovanna, hoping that she might still have the diaries' originals in her possession. I will tell of this expedition's results later on in this preface.

# Zur Quantenmechanik des radioaktiven Kerns.

Von **G. Gamow** z. Z. in Kopenhagen und **F. G. Houtermans** in Berlin.

Auf Grund der von G. Gamow gegebenen quantenmechanischen Deutung des $\alpha$-Zerfalls werden die Zerfallskonstanten der $\alpha$-Strahler aus der Energie der $\alpha$-Teilchen, den Atomnummern und gewissen charakteristischen Kernradien berechnet. Für je ein Element der radioaktiven Familien wird der Kernradius so berechnet, daß die Zerfallskonstante mit der Erfahrung übereinstimmt. Daraus werden — ohne spezielle Annahmen über den Potentialverlauf in unmittelbarer Nähe des Kerns — die Zerfallskonstanten der übrigen $\alpha$-Strahler berechnet und in genäherter Übereinstimmung mit den experimentellen Werten gefunden. Daran schließen sich einige qualitative Betrachtungen über den Mechanismus des radioaktiven Zerfalls.

In einer kürzlich erschienenen Arbeit* hat der eine von uns (G. G.) allgemeine Gesichtspunkte für das quantenmechanische Verständnis des $\alpha$-Zerfalls radioaktiver Elemente angegeben. Die angenäherte Rechnung ergab dort schon qualitativ den der Geiger-Nutallbeziehung zugrunde liegenden Zusammenhang zwischen der Energie der $\alpha$-Teilchen und der Zerfallskonstante der $\alpha$-Strahler, auch ließ sich danach theoretisch für den Fall des Ra C′ eine experimentell gefundene** Abweichung von der Geiger-Nutallbeziehung erwarten.

Hier soll nun versucht werden, die absolute Größe der Zerfallskonstanten durch eine genauere Rechnung zu finden und mit den experimentellen Daten zu vergleichen. Dabei ergibt sich die Möglichkeit, einige weitere Folgerungen über die Struktur radioaktiver Kerne zu erhalten.

§ 1. Wir betrachten den Kern als Ausgangspunkt eines gedämpften $\psi$-Wellenzuges. Die ausgeschleuderten $\alpha$-Partikel entsprechen der weglaufenden $\psi$-Welle, die Verminderung der $\psi\overline{\psi}$-Menge des $\alpha$-Teilchens im Kern bedeutet den radioaktiven Zerfall.

Betrachten wir ein positives Teilchen, das sich unter dem Einfluß einer Zentralkraft um einen positiven Kern bewegt und eine positive Gesamtenergie $E$ hat. Dann lautet die Schrödingersche Amplitudengleichung unter vorläufiger Vernachlässigung der Dämpfung

$$\frac{\partial^2 \Psi}{\partial r^2} + \frac{2}{r}\frac{\partial \Psi}{\partial r} + \frac{8\pi^2 m}{h^2}(E - U)\Psi = 0, \qquad (1)$$

* G. Gamow, ZS. f. Phys. **51**, 204, 1928. Vgl. auch E. Condon und J. C. Gurney (Nature **122**, 439, Sept. 1928), die genau die gleichen Gesichtspunkte wie die zitierte Arbeit zur Erklärung des $\alpha$-Zerfalls heranzieht.

** J. C. Jacobsen, Phil. Mag. **47**, 23, 1924.

The first page of the Gamow-Houtermans paper on decays of radioactive atoms from *Zeitschrift für Physik*, 52, 496 (1928).

— 3 —

The reader may now find it helpful to have a sketch of the setting for the events to be narrated, as well as of the key players in this drama.

The Kharkov Fiztech (formerly the Ukrainian Physico-Technical Institute, UPTI) was founded by Academician Abram Ioffe in 1928. In the early 1930s and until the Great Purge in 1937-38, it was the leading physics research center in the USSR. In 1932 Lev Landau became the head of the Theory Department of the Institute, and by 1937, when Landau had to flee Kharkov, his department had become a top-level group. UPTI was the site of the first Soviet experimental splitting of an atomic nucleus (so-called induced fission). In 1940, just a year before the German invasion of the USSR, a memo written in this institute by Friedrich Lange, Vladimir Shpinel and Victor Maslov proposed making an A-bomb based on nuclear fission. In fact, these authors invented the concept of the atomic bomb, as well as a method of separation of the Uranium-235 isotope from uranium ore. Who knows how history might have turned if their proposal, "The Use of Uranium as an Explosive," had been set in motion at the time.

The scientific organization of the UPTI was as follows [3]. "There was an experimental low-temperature group under Lev Schubnikov, Boris Lasarev, and Abram Kikoin; there was also a low-temperature group with Martin and Barbara Ruhemann oriented toward industrial application. By the way, the Ruhemanns wrote a fine book on low-temperature physics, one of the first along this line. It was intended that this group should become an independent institute. The idea of this institute was conceived by Alexander Weissberg, a Viennese engineer and a communist. He joined the UPTI in the early 1930s. He convinced the Commissariat of Heavy Industry of the soundness of applied cryogenics, and he was commissioned to build the new institute. During my May 1934 visit I was enormously impressed by his story, and during my entire stay there beginning in January 1935 he acted as a contractor.

I considered it impressive that the apparently rigid system could act with such flexibility. However, all ended differently... Furthermore, an electronics group was working on secret radar problems, and a successful neutron-physics group was established by Fritz Houtermans, a German communist. Last but not least, there was the theory group, headed by Landau.

Landau reviewed the physics journals each week. There was an excellent research library, and he assigned papers to be reported on in the seminars, three-to-four papers for each session... His judgement was accepted without any questions.

The UPTI campus was at the edge of the city. It contained a dormitory, a cafeteria, three-room apartments for the senior staff, and a tennis court, apart from the laboratories and workshops.

There were occasional parties of dancing and singing, involving both the scientists and the technicians of the UPTI."

UPTI's first director, appointed in 1929, was Ivan Obreimov. When Obreimov was charged with the task of turning UPTI into a world-class research center, he came up with a brilliant idea: to invite to Kharkov eminent German and Austrian physicists whose situation under Hitler was critical because they were either communist sympathizers, or Jews (sometimes both inconvenient qualities coincided in one person). The first Western physicist hired by Obreimov was Walter Elsasser.[6] In his memoir [4] Elsasser described his conversation with Obreimov and and his reasons for accepting the offer.

"Some time late in 1929 the telephone rang: `This is Obreimov. I am in Berlin and would like to see you.' Obreimov was an experimental physicist who had been in Leiden at the same time as I; and since we stayed at the same rooming house, we had become well acquainted. He told me that he had been made director of the

[6] Walter Maurice Elsasser (1904-1991), a German-American physicist, a "father" of the presently accepted dynamo theory explaning the Earth's magnetism. He proposed that the Earth's magnetic field resulted from electric currents induced in the fluid outer core of the Earth.

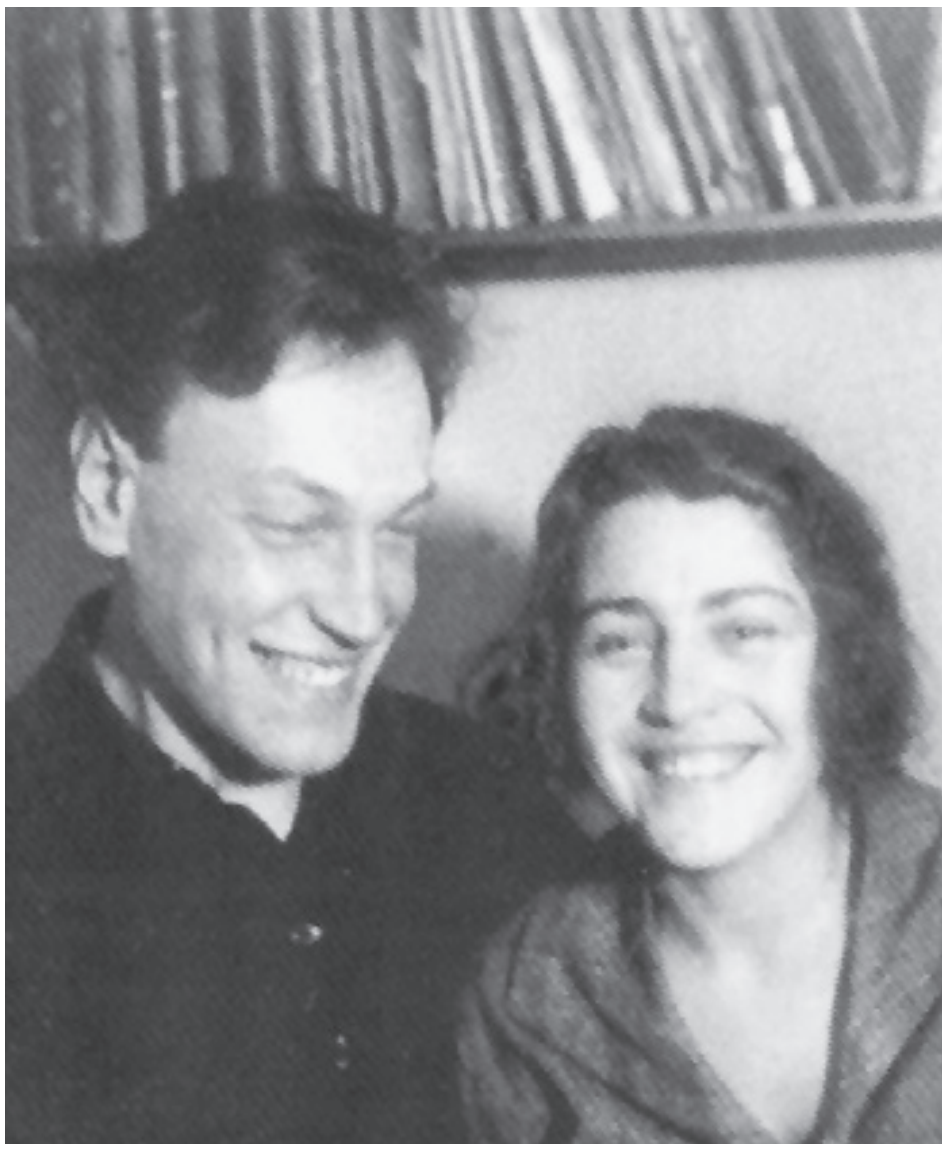

**Alexander Ilyich Leipunsky with his wife Antonina Prikhotko**

Ukrainian Physico-Technical Institute in Kharkov, a large industrial town in the Ukraine. … Would I be interested in coming to Kharkov for a year as a `technical specialist' under a suitable contract?.. Half of my salary would be paid in rubles that could be not taken out of Russia, the other half in marks or any other currency convertible on the world market. The sum he mentioned would have been generous for a reasonably experienced practical engineer; for me it was princely. He also informed me that I was the first non-Russian to be associated with the Institute, so it was a thoroughly experimental and challenging undertaking … After a short hesitation I agreed. Although it might not be beneficial to my career as a scientist — and ultimately it wasn't — it offered both a new possibility of escaping from Germany and a great adventure."

Elsasser found UPTI little more than a glorified construction site and cut short his stay [5]. Later Obreimov's idea was fully implemented by Alexander Leipunsky[7], who succeeded Obreimov as

[7] Alexander Ilyich Leipunsky (1903-1972), was born in Poland (although in 1903 his native village was a part of the Russian Empire) into a Jewish family and received his education in Leningrad under Academician Abram Ioffe. In 1937 he was purged from the Communist Party "for aiding the enemies of the people" and removed from his post as UPTI Director. Arrested in Kharkov, Leipunsky was kept in a Kiev prison for two months. Neither A. Weissberg nor F. Houtermans gave any incriminating evidence against him. Leipunsky's case was closed and he was released, an incredibly rare — almost unheard of — occurrence in

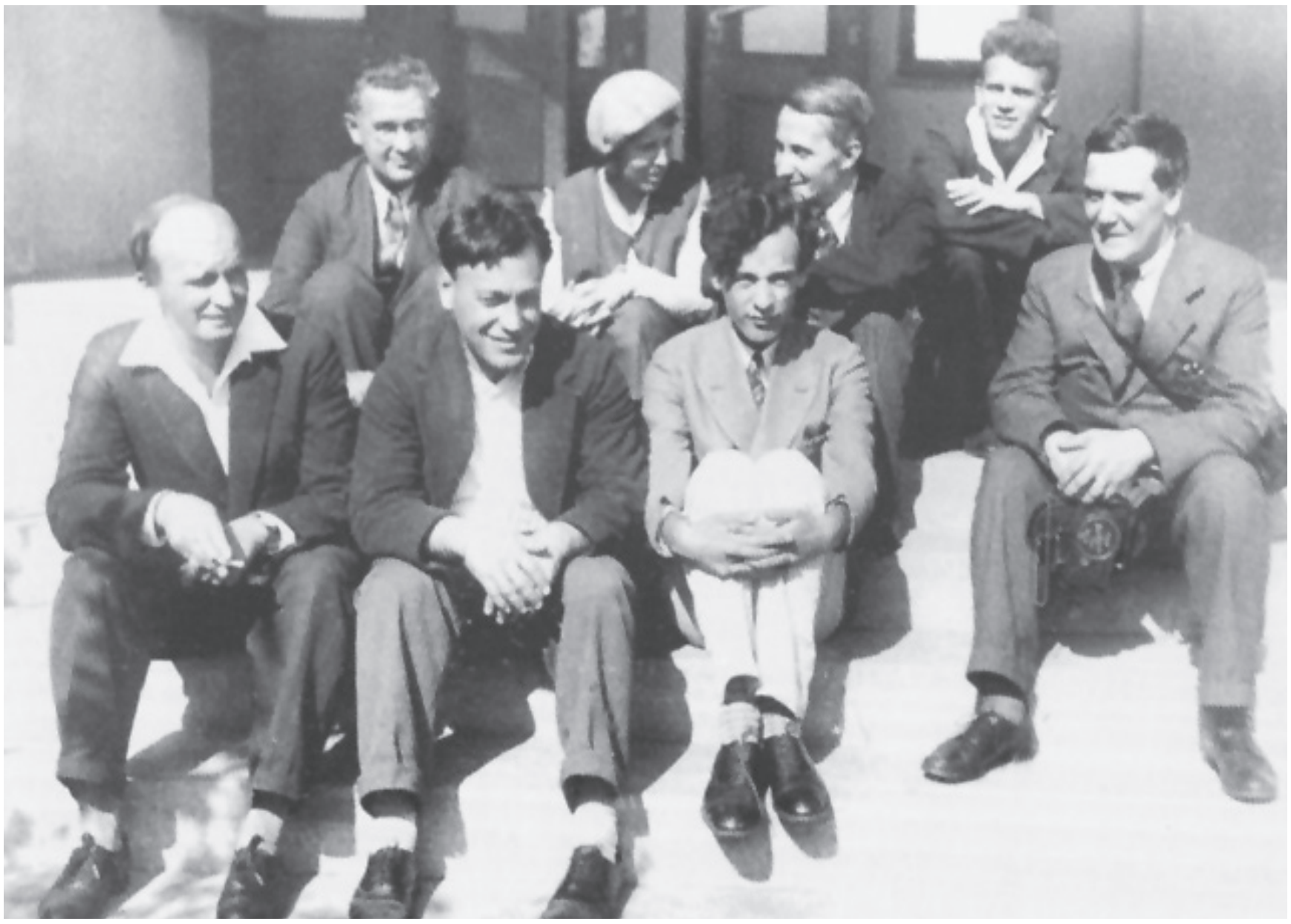

**Alexander Leipunsky with his UPTI colleagues in September 1934 at the entrance to the Main building of UPTI. In the first row from left to right: Lev Shubnikov, Alexander Leipunsky, Lev Landau, Pyotr Kapitsa.**

UPTI's director in 1933.

— 4 —

I hold in my hands a book by Alexander Weissberg, entitled *The Accused* [6]. It was published in 1951 in New York and, in fact, was one of the first testimonies of a living witness and a victim of the Soviet Great Terror published in the West. I stress, the year of publication was 1951, long before Solzhenitsyn's *Gulag Archipelago*. Let us remember this name, Alexander Weissberg: we will encounter it more than once in the pages of this collection. The

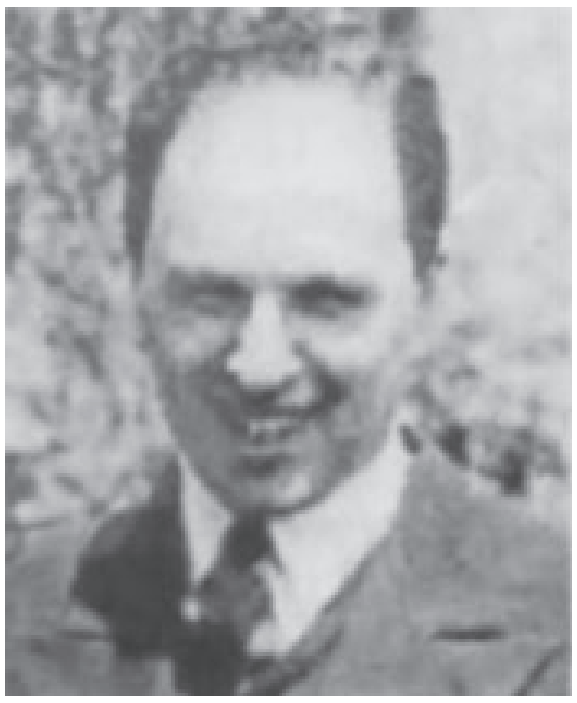

**Alexander Weissberg**

judicial practice in the Soviet Union.

foreword to Weissberg's book was written by Arthur Koestler, the famous author of *Darkness at Noon* [7].

Weissberg writes in Chapter 1:

> The aim of this book is to describe happenings without precedent in modern history. From the middle of 1936 to the end of 1938 the totalitarian state took on its final form in the Soviet Union. In this period approximately 8 million people were arrested in town and country by the secret police (NKVD at that time). The arrested men were charged with high treason, espionage, sabotage, preparations for armed insurrection and the planning of attempts on the lives of Soviet leaders. ... All these men, with very few exceptions, pleaded guilty.
>
> They were all innocent...

Then he continues, with bitterness:

> I know that I shall be fiercely assailed by those [in the West] who have made it their business to defend the system of totalitarian lies. I know that like all others who have come forward in the past I shall be ruthlessly slandered. I cannot prevent that...

— 5 —

Alexander Leipunsky was a charismatic person. People at the Institute not only respected but loved him. Edna Cooper (K. Sinelnikov's[8] wife), whom we will meet in the pages of this book, writes [8]: "Alexander Leipunsky, was, as always, charming, it seems that all women in the Institute were in love with him. I'm trying to be an exception." Charlotte Houtermans notes in her diary: "He was *sympatico*, gentle, entertaining, pleasant and intelligent."

[8] Kirill Dmitrievich Sinelnikov (1901-1966), an experimental physicist who was the first to carry out lithium fission in 1932. At UPTI from 1930. Edna Cooper (1904-1967) and Kirill Sinelnikov were married in Cambridge, UK, in 1930, shortly before Sinelnikov's return to the USSR.

In Chapter 2, Weissberg writes:

> Communist Leipunsky was appointed as "Red Director." ... His way of life was extremely modest. In 1931 I went with him to Moscow to obtain permission for the publication of a journal of physics. I had been in the Soviet Union only a few weeks, but he and I already got on excellently... Before we went, I noticed the state of Leipunsky's boots. All my Western ideas of propriety revolted at appearing with broken boots for an interview with a high government official.
>
> "Alexander Ilyich," I declared reproachfully, "you can't possibly go to see the People's Commissar with these boots. After all, you're Director of the Institute."
>
> "So what?" he inquired. "They are the only ones I've got and I can't afford to buy any more, with things the prices they are now[9]."

The journal mentioned above was *Physikalische Zeitschrift der Sowjetunion*, which during the six years of its existence — from 1933 till 1938 — played a very important role in connecting physicists from the East and the West. Papers by major players on the theoretical scene, including such titans as P. Dirac, L. Landau, and M. Bronshtein, were published in Russian, German and English in this journal. After UPTI's fall, such multilingual publications resumed in Russia only after the collapse of the Soviet Union in 1991.

On June 14, 1938, Alexander Leipunsky was arrested by NKVD as a Polish spy. He spent exactly eight weeks in prison under almost continuous interrogation, and left the NKVD prison on August 9 a

[9] According to Weissberg, Leipunsky's monthly salary at that time was 280 rubles. The price of a pair of shoes was 100-120 rubles.

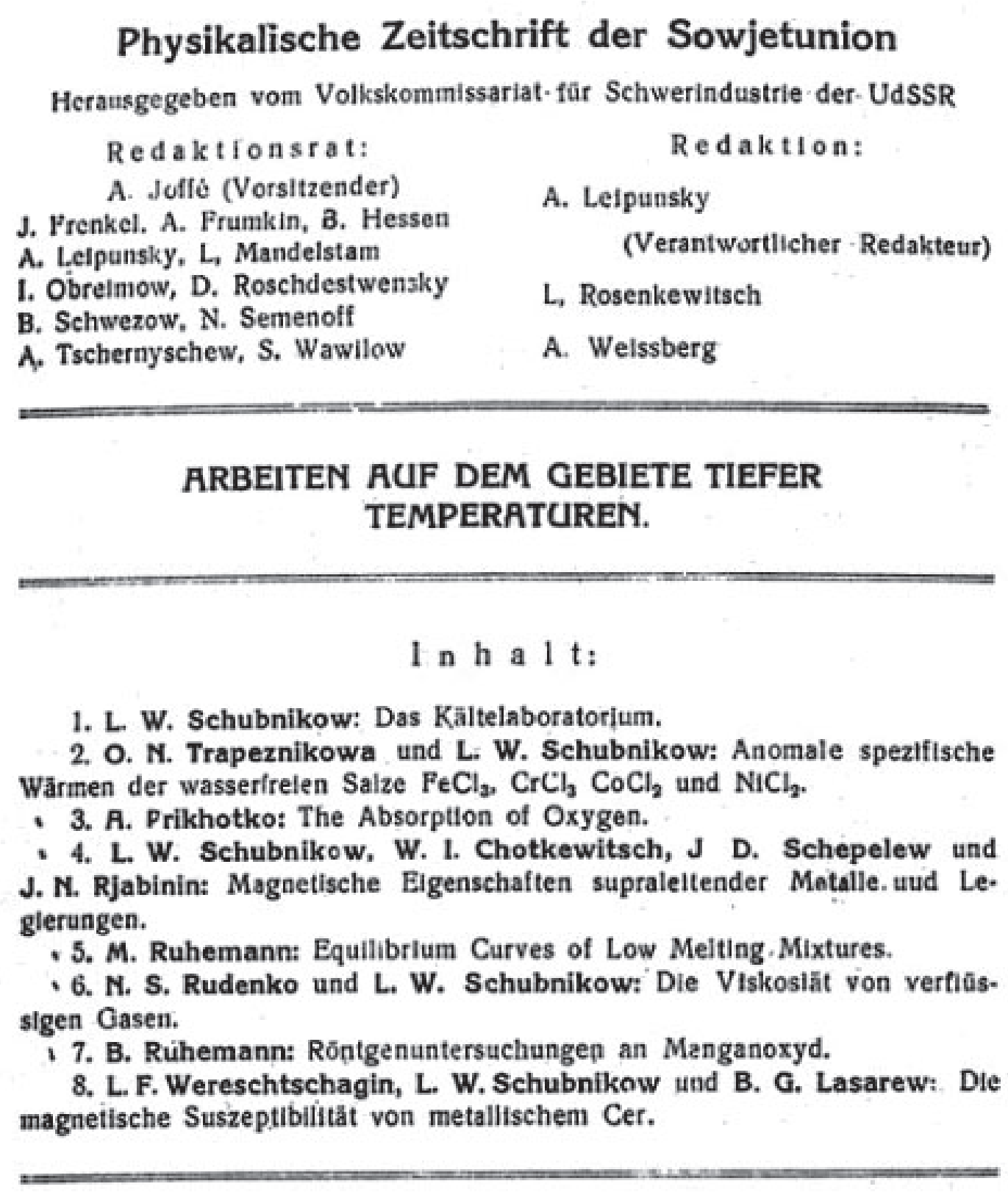

**Physikalische Zeitschrift der Sowjetunion**

Herausgegeben vom Volkskommissariat für Schwerindustrie der UdSSR

Redaktionsrat:

A. Joffé (Vorsitzender)
J. Frenkel, A. Frumkin, B. Hessen
A. Leipunsky, L. Mandelstam
I. Obreimow, D. Roschdestwensky
B. Schwezow, N. Semenoff
A. Tschernyschew, S. Wawilow

Redaktion:

A. Leipunsky
(Verantwortlicher Redakteur)
L. Rosenkewitsch
A. Weissberg

**ARBEITEN AUF DEM GEBIETE TIEFER TEMPERATUREN.**

Inhalt:

1. **L. W. Schubnikow**: Das Kältelaboratorium.
2. **O. N. Trapeznikowa** und **L. W. Schubnikow**: Anomale spezifische Wärmen der wasserfreien Salze $FeCl_3$, $CrCl_3$ $CoCl_2$ und $NiCl_2$.
3. **A. Prikhotko**: The Absorption of Oxygen.
4. **L. W. Schubnikow, W. I. Chotkewitsch, J. D. Schepelew** und **J. N. Rjabinin**: Magnetische Eigenschaften supraleitender Metalle uud Legierungen.
5. **M. Ruhemann**: Equilibrium Curves of Low Melting Mixtures.
6. **N. S. Rudenko** und **L. W. Schubnikow**: Die Viskosität von verflüssigen Gasen.
7. **B. Ruhemann**: Röntgenuntersuchungen an Manganoxyd.
8. **L. F. Wereschtschagin, L. W. Schubnikow** und **B. G. Lasarew**: Die magnetische Suszeptibilität von metallischem Cer.

**The titlepage of a volume of Phisykalische *Zeitschrift der Sowjetunion* in 1936. This journal was published by the People's Commisariat of Heavy Machine Building. That's where Weissberg and Leipunsky were heading.**

different man, completely broken.

The NKVD extorted the following confession [9] from Leipunsky:

> My work as Director of UPTI was extremely detrimental to the development of Soviet science. Although I was subjectively not associated with enemy operatives at the Institute, my activity objectively aided the enemy. Only through my assistance and support was the enemy able to sabotage UPTI so prolongedly. My support of the enemy consisted of the following:

> 1. I brought the spy Houtermans to the USSR and created favorable conditions for his espionage work.
>
> 2. I covered up signs of the spy Weissberg's hostile intentions, provided him with various forms of assistance, and tried to keep him at the institute; I temporarily hid from the NKVD the fact of his obtaining information through espionage, and gave him a positive employee appraisal at his dismissal from the institute.
>
> 3. I covered up signs of the hostile intentions of the counter-revolutionaries Landau and Shubnikov, and tried my best to keep them at the institute. I created conditions at the institute for their engagement in sabotage and espionage work. All this happened because I, as a result of my total political recklessness and rotten liberalism, overestimated the importance of connections with Western European science, played lackey to the West, and made conditions at the institute exceptionally favorable for the enemy.
>
> Leipunsky

— 6 —

Since both Leipunsky and Weissberg played such a fateful role in the life of F. Houtermans, I'd also like to acquaint the reader further with Alexander Weissberg. In 1931 Weissberg (or Weissberg-Cybulski, as he is referred to in some documents) became the first foreign physicist hired to work at UPTI on a permanent basis. This was a relatively good time in the USSR. Alexander quickly made friends, who started calling him Alexander Semyonovich, following traditional Russian usage of a first name and patronymic. He knew many Western physicists, and he was supposed to entice to Kharkov the most prominent of those who potentially might accept.

Here is what Koestler writes in his preface to *The Accused*.

> Alex Weissberg was born in 1901 in Krakow, which then

> belonged to Austria. His father was a prosperous Jewish merchant. A few years after his birth his family moved to Vienna... He studied mathematical physics and engineering in Vienna, where he was graduated in 1926. In 1927 he joined the Communist Party.
>
> In 1931 he received an offer from the UPTI in Kharkov and decided to move to the Soviet Union for good...
>
> In Kharkov I stayed in the Weissbergs' flat. It was a small but by Russian standards luxurious flat in the vicinity of the Institute. The latter was one of the largest and the best-equipped experimental laboratories in Europe. During my stay with Alex and his wife, I met most of the scientists who appear as *dramatic personae* in this book. Among them were Leipunsky and Landau, the infant prodigy of Russian physics... I remember a long discussion with Landau, who argued with great conviction that the works of all the philosophers from the beginning of time up to and excluding Marx are not worth the paper on which they are printed.

Among other things, Koestler also tells the story of Alexander Weissberg's personal life. Before emigrating to the USSR Alexander Weissberg was engaged to Éva Striker, an artist and ceramics designer of Jewish-Hungarian descent.[10] Weissberg arrived in Kharkov in 1931 with a firm intention to establish himself as a part of a paradise state of workers and peasants. Éva later joined him in Kharkov, where they were married. Their marriage lasted only a few years, ending in separation in 1934 when she left Kharkov for Moscow. On May 26, 1936 Éva was arrested by the NKVD. One of the charges brought against her was

[10] In 1938 after her second marriage in England Éva became Eva Zeisel. After subsequent emigration to the US she eventually made her way to the summit of the artistic Olympus and became an internationally acclaimed designer. Eva Zeisel received many distinguished awards, e.g. the 2002 Living Legend Award from the Pratt Institute, the Middle Cross of the Order of Merit award of the Republic of Hungary (2004), the 2005 Cooper-Hewitt National Design Award for Lifetime Achievements. Eva Zeisel died in 2011 at the age of 105.

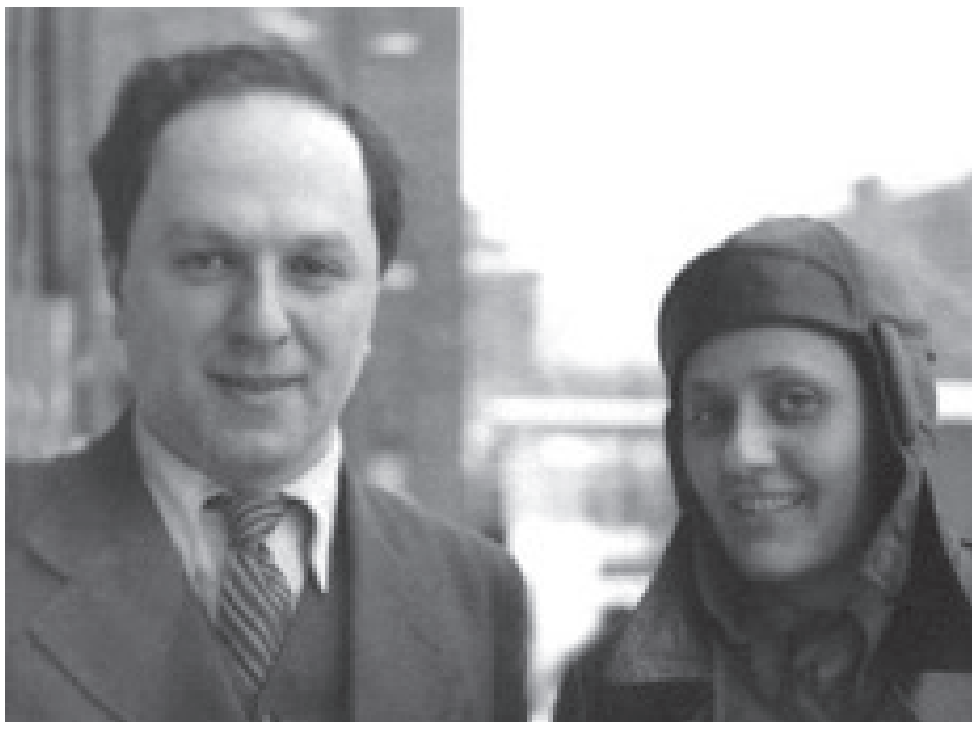

Alexander and Éva. Reproduction by courtesy of Jean Richards.

her alleged patrticipation in a plot to assassinate Stalin. After a few days of imprisonment in Moscow's Butyrki, she was transferred to a Leningrad prison where she was held for 16 months, 12 of them in solitary confinement.

Learning of Éva's arrest, Alexander Weissberg at once realized the probability of his own. Koestler writes that because Weissberg had an Austrian passport, and it was only 1936 on the calendar, he could have gotten an exit visa from the USSR had he acted quickly and, in particular, appealed for help to the Austrian Embassy (still two years before the Anschlüss!). Instead, Alexander rushed to Moscow to solicit for Éva. He appealed to higher authorities with whom he had dealt previously in connection with a high-profile gas liquefying facility which he oversaw at UPTI. He even managed to get an audience with USSR Prosecutor General Andrei Vyshinsky. In September of 1937 Éva was deported to Austria. But for Weissberg himself it was too late.

The story of Weissberg's arrest by the NKVD and imprisonment is described in Frenkel's book and, in more detail, in [6]. In 1940, after the Hitler-Stalin pact, the NKVD handed over to the German Gestapo a group of Austrian and German communists and Jews. Weissberg was in this group. His subsequent life is no less remarkable

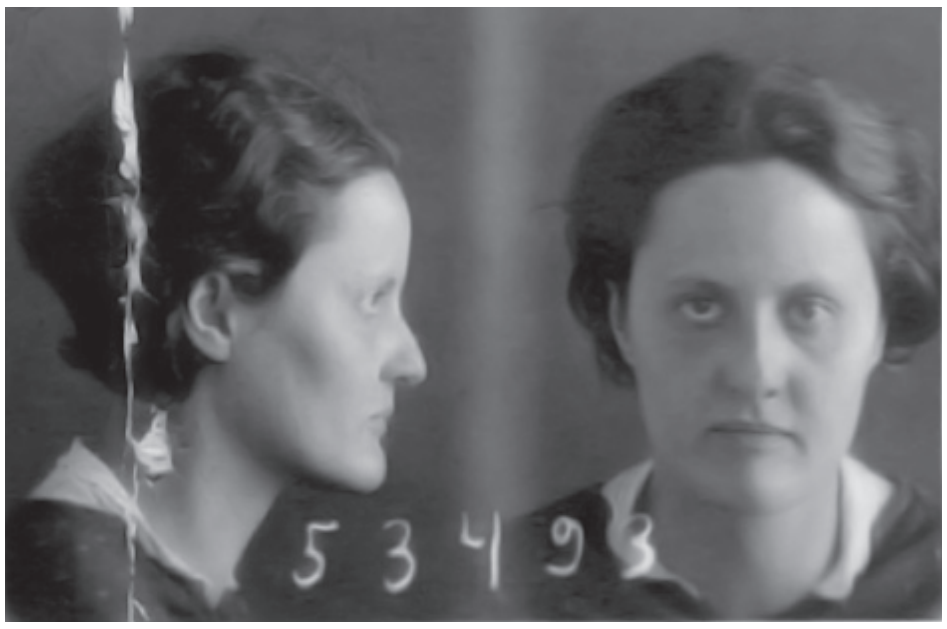

An NKVD prison photograph of Éva Striker.

[10]. The Germans sent the former NKVD prisoners first to a prison in Biala Podlaska, then moved them to Warsaw's Pawiak prison, where they conducted racial selection. Five Jews, including Weissberg, were transferred to the prison at Lublin Castle; others were transported to concentration camps in the Reich.

After a two-month stay in the Lublin prison Weissberg was sent to the Krakow ghetto with a Star of David armband on his arm. "This sealed my fate as a physicist. After many years on the run I could not go back to my former vocation," he would later write.

After finding out that he was on a German list of the ghetto dwellers singled out for execution, Weissberg escaped from Krakow and stayed in the ghettos in Bochnia and Tarnow until September 1942. At the beginning of the mass extermination of the Polish Jews he went underground and moved to the "Aryan side" of Warsaw. Here he hid in the apartment of Zofia Cybulska, his future wife. On March 4 of 1943, Weissberg was arrested by Gestapo agents in her apartment. Zofia Cybulska had to go into hiding immediately. Weissberg was sent to the Pawiak prison again, and then to a concentration camp in Kaweczyn. He escaped from there with the help of a German foreman. In Warsaw, Zofia Cybulska again took care of him. They survived the Warsaw uprising, in which Weissberg took an active part. After the defeat of the uprising a German saved him from the Pruszkow camp. Weissberg spent the last months of the German occupation in hiding in Wlochach near Warsaw. When he married Zofia Cybulska he added his wife's name to his, seeking to conceal his identity in anticipation of persecution by the NKVD, which came to Poland with the Red Army.

In mid-1946 he managed to escape from Poland to Sweden, where he was soon joined by his wife. Later the family settled in Paris.

In the 1950s Michael Polanyi enlisted Weissberg as a participant in the work of the Congress for Cultural Freedom, an anti-communist advocacy group founded in 1950. He died in 1964

— 7 —

In 1932 a man-made disaster known as *Holodomor*, or Extermination by Hunger, struck Ukraine. It was caused by Stalin's ruthless campaign of so-called farm *collectivization*. The property of the best farmers (considered to be enemies of socialism) was expropriated, and they themselves were exiled to Siberia with no means of survival. The famine that ensued in Ukraine was so severe that in 1932 and 1933 several million people died from starvation. Orders were given to shoot starving people should they try to escape the zone of famine. Some idea of life in Kharkov in these years can be inferred from snapshots made by an Austrian engineer, Alexander Wienerberg, who spent 19 years in Russia, until 1934, and who witnessed all these events. As of 1933 he was a technical director at a synthetic chemical factory in Kharkov. For me it is hard to understand how Weissberg, and other Westerners who arrived at UPTI later on, could have failed to notice this disaster.

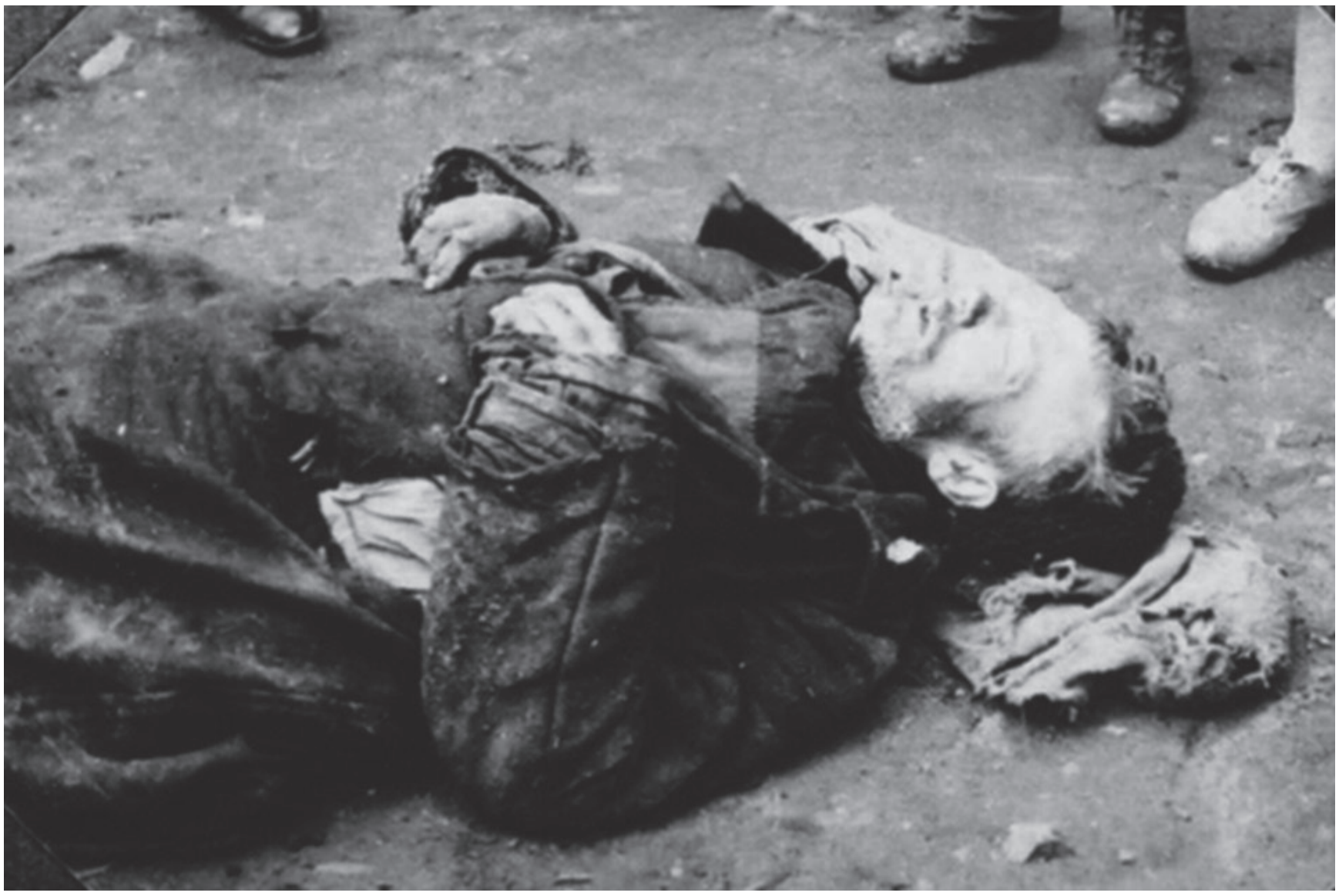

**The corpses of the starved in the streets of Kharkov at first aroused sympathy.**

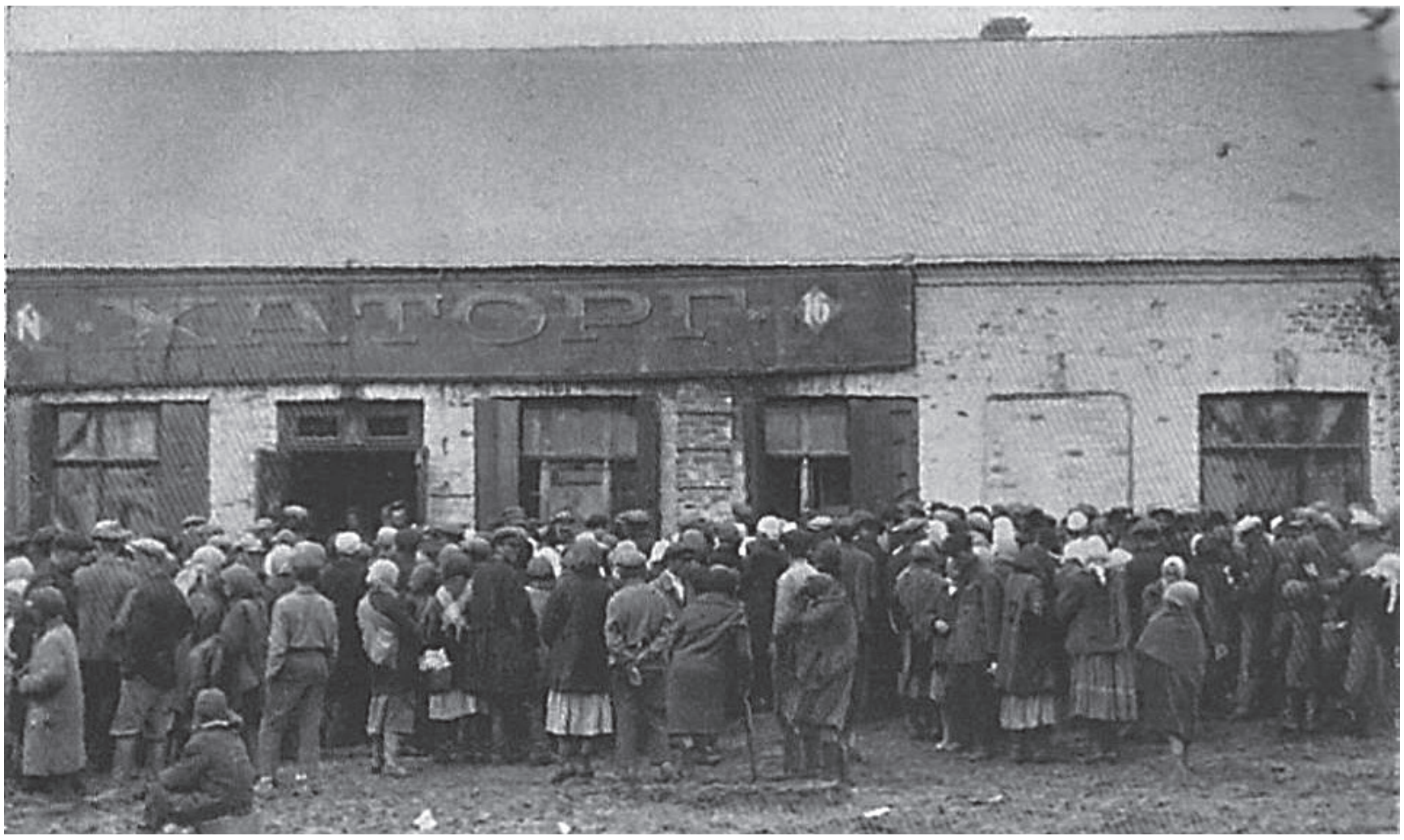

**The empty "Khartog" (Kharkov Trade Cooperative) food distribution site is besieged by a devastated population in Kharkov.**

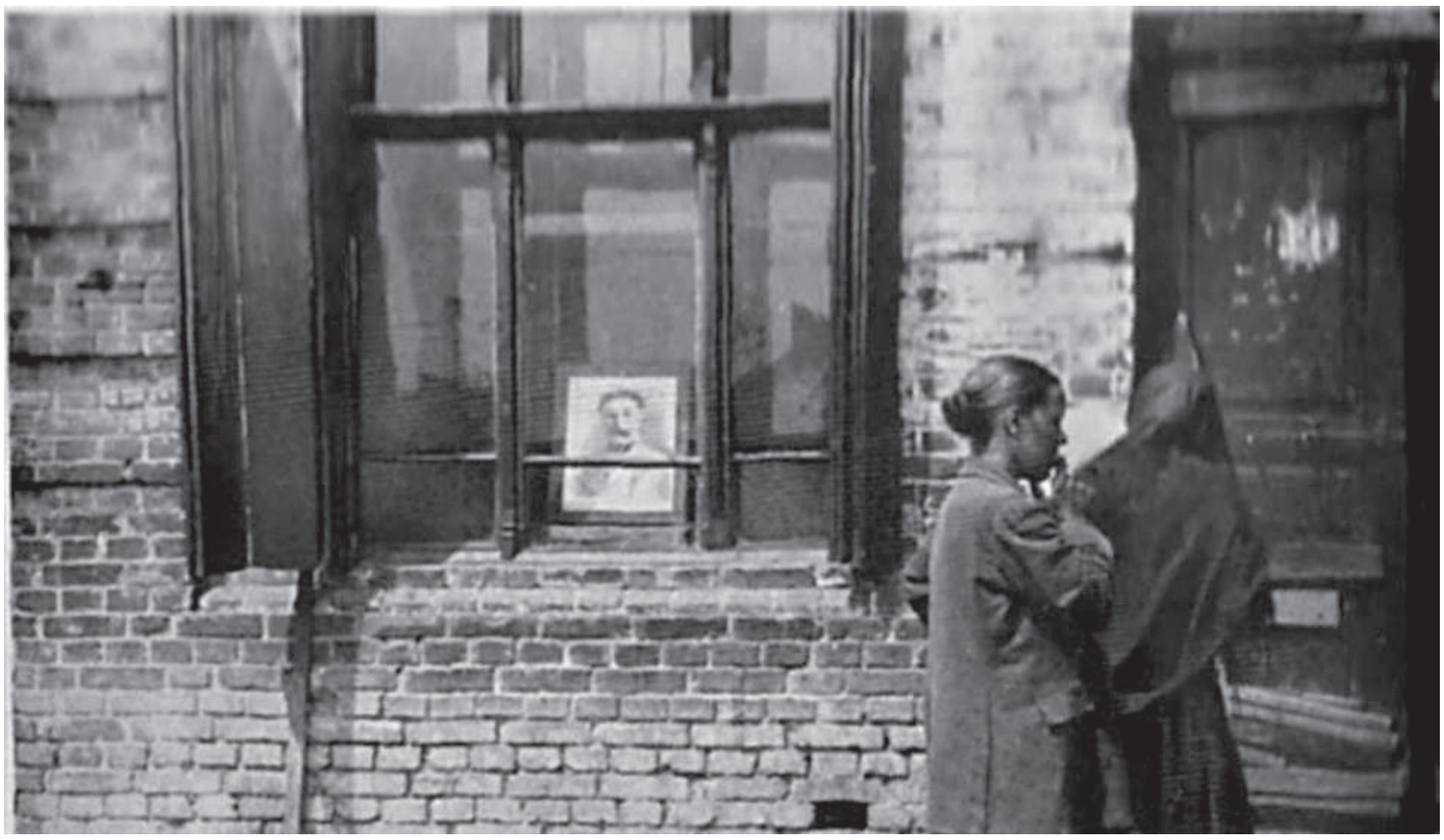

**The windows of the empty food places decorated only by pictures of Stalin and other Muscovite rulers.**

Of course, at that time Weissberg's vision could have been blurred by an unshakable belief in the communist society's radiant future.[11] Yet it wouldn't be long before he changed his mind.

— 8 —

*Holodomor* did not have such an immediate and devastating impact on UPTI as on the area's general populace, since UPTI was considered of paramount importance to the Soviet government, with food supplies distributed among the UPTI employees. The foreign members of the group were additionally privileged because a part of their salaries was paid in hard currency, which opened them the doors of special stores closed for ordinary Soviet citizens.

The advent of the Great Purge was imminent. A careful and unbiased observer could feel it in the air. The first to be arrested on political charges, in the spring of 1936, was Éva Striker, Alexander Weissberg's wife. In 1937, when the Great Purge was fully unleashed, UPTI was decimated, with virtually all leading researchers ending up in the Gulag, executed or deported. Among those sentenced to death were Lev Shubnikov (1901-1937), Lev Rozenkevich (1905-1937), Vadim Gorsky (1905-1937), Valentin Fomin (1909-1937), and Konrad Weisselberg (1905-1937). After the first arrests, Lev Landau, the future Nobel Prize winner, who headed the UPTI theory department, left for Moscow but was arrested there in April 1938. He spent a year in NKVD prison, and only Pyotr Kapitsa's appeal to Stalin saved his life. Moisei Koretz (we will encounter his

[11] In Weissberg's book (p. 212) I found this illuminating passage: "We all knew the truth, but we were all convinced that socialism would be victorious in the end. We knew that the famine was not an act of God, but due to Stalin's false policy, and we hoped he would soon see his mistake and correct it. Not one of us even thought of overthrowing him, or even calling him to account. That would have been impossible without a political revolution, and anything of the sort would have meant the victory of the White counterrevolution supported by masses of starving peasants."

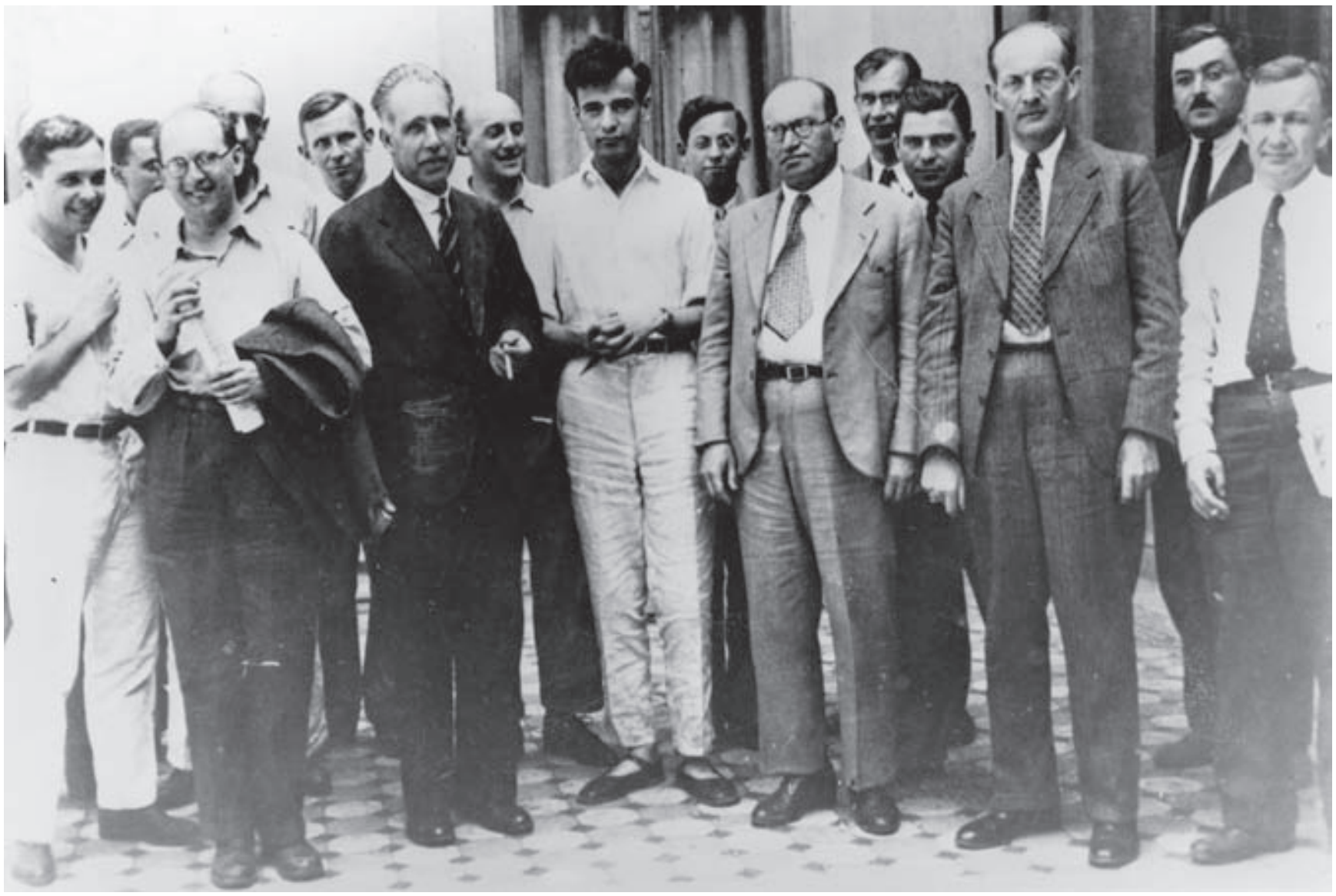

**Participants in the meeting at the UPTI in Kharkov, May 1934. Left to right: D. Ivanenko, L. Tisza (obscured), L. Rosenfeld, unknown (obscured), Yu. Rumer, Niels Bohr, J. D. Crowther, L. Landau, Milton Plesset, Yakov Frenkel, Ivar Waller, E. J. Williams, Walter Gordon, V. A. Fock, I. E. Tamm. Credit: American Institute of Physics Emilio Segrè Visual Archives, *Physics Today* Collection.**

name more than once in the second part of the book), a colleague and collaborator of Lev Landau, suffered a similar fate: arrested in 1938 as Landau's co-conspirator, Koretz spent 14 years in the Gulag. Altogether 11 UPTI employees were run down by Stalin's repression machine [11].

UPTI ceased to exist as a center of excellence for theoretical and experimental physics. Its happy golden years were gone. On September 1937 at the Second All-Union conference on nuclear physics in Moscow, the participants in the conference addressed Comrade Stalin with these passionate words of admiration:

"The successful development of Soviet physics occurs against the background of a general decline of science in capitalist countries, where science is falsified and is placed at the service of greater exploitation of man by man... Vile agents of fascism, Trotsky-ist-Bukharinist spies and saboteurs ... do not stop short of any abomination to undermine the power of our country ... Enemies penetrated among physicists, carrying out espionage and sabotage assignments in our research institutes ... Along with all the working people of our socialist motherland, Soviet physicists more closely unite around the Communist party and Soviet government, around our great leader Comrade Stalin ..."

A special apartment house built by UPTI in 1928 as a residence for its most distinguished members, a house which saw such guests as future Nobel laureates Niels Bohr, Werner Heisenberg, Paul Dirac, John Cockcroft, Pyotr Kapitsa, Nikolay Semenov, Irene and Frederic Joliot-Curie, and Igor Tamm, fell into vacancy...

— 9 —

Alexander Weissberg, after his arrival to Kharkov in 1931, attracted a small constellation of other top foreign scientists. In the summer of 1931, Weissberg was back in Vienna on holiday, with the mission to find and hire experimentalists. "Friends recommended Martin Ruhemann.  The British-born Ruhemann moved to Germany as a young man. His German wife, Barbara, was also a physicist...  Ruhemann worked in the low-temperature laboratory and in 1935 he and Weissberg proposed a plan for a new low-temperature research station, to develop the local nitrogen industry and other branches of the Soviet chemical industry [5]."

Between 1935 and 1937, Ruhemann was the prospective scientific director of this large, new development, with Weissberg as director and manager of the construction. Weissberg's description of Ruhemann's initial reaction to the living and working conditions in Kharkov, followed by his gradual change of heart, is illuminating:

"When he first came to the Soviet Union in 1932 the country was experiencing the worst year since the end of the civil war. At first he and his wife were always on the point of leaving. All they could see was chaos, poverty and hunger. … He had never thought much about the significance of social revolutions but once in the Soviet Union he grasped what was really at stake and was swept up in a new and larger movement than he had ever known before [6]."

Another Western arrival in 1935, the Hungarian László Tisza, who was trained in Göttingen and Leipzig, represented a younger generation.

Here is an excerpt from a 1965 interview,[12] with Victor Weisskopf, who went on after Kharkov to serve as the head of a theoretical physics group at Los Alamos, New Mexico, during the Manhattan Project, and, later, as Director General of the European Center for Nuclear Research (CERN) at Geneva:

> I went to Russia for one-half year, from February until my Rockefeller [fellowship] started, which was October or November 1932. I was there altogether about nine months. They invited me. I had a number of friends in Kharkov; Viennese who went there, communists who moved over. It was a very bad time. A lot of people who were either communists, half-communists or non-communists who just went there because it was the only place where you could stay alive. There was a new institute in Kharkov; Landau was there, as was Alexander Weissberg... He is a converted man now. He is a very interesting man: mostly a politician but started off in physics. He, at that time, was a big shot in Kharkov; he invited me. Placzek[13], also came and Houter-

[12] This and Tisza's interview below are quoted according to [12].

[13]Georg Placzek (1905-1955) was a Czech physicist, born in Brno, Moravia to Jewish parents. Placzek studied physics in Prague and Vienna. He worked with Hans Bethe, Edward Teller, Rudolf Peierls, Werner Heisenberg, Victor Weisskopf, Enrico Fermi, Niels Bohr, Lev Landau, Edoardo Amaldi, Emilio Segré, Leon van Hove and many other prominent physicists of his time, in the areas of Raman scattering, molecular spectroscopy in gases and liquids, neutron physics and mathematical physics. Together with Otto Frisch, he suggested a direct experimental proof of nuclear fission. Together with Niels Bohr and others, he was instrumental in clarifying the role of Uranium-235 for the possibility of nuclear chain reaction. During his stay in Landau's circle in Kharkov, Placzek witnessed the brutal reality of Stalin's regime. His first-hand experience of this influenced the political opinions of his close friends, Robert Oppenheimer and Edward Teller in particular. Later, Placzek was the only Czech with a leading position in

> mans was there.
>
> Houtermans was an assistant of Richard Becker during the Berlin years, and I was very much in contact with him there. At Houtermans' house in Berlin I kept up in physics, but mostly I was busy with politics and met many people of the leftist-liberal type. That was an exciting time in Berlin. At any rate, Houtermans was later in Kharkov, which was sort of a receptacle of refugees, either depression refugees or Nazi refugees... László Tisza was also there.

The group of western expats at UPTI included, among others, Fritz Lange, Konrad Weisselberg, Ürgen Peters (Lange's assistant), etc. I will say a few words about them, starting with the Ruhemanns.

— 10 —

When Ruhemann's friends began to disappear one by one, and his contract was not renewed, he applied for exit visas. In his unpublished book *Half a Life*, Ruhemann wrote that he did not betray the ideals of communism, but realized that even his ideological purity was no guarantee against Stalin's purges. And further:

> In 1937, we began to move away from our Kharkov environment. It was very easy, because all of our friends and colleagues had already moved away from us, treating us as dangerous foreigners.

After a six-month nerve-wracking wait, exit visas were received and the Ruhemanns family made it safely to England, where husband and wife lived long lives together (Martin died in 1994) doing research in things they loved — cryogenics and gas separation. Yet it's important to realize that the toxic atmosphere of a state where

---

the Manhattan Project.

the government tells each and everyone of its citizens what is right and what is wrong, the pressure of fear, can break even decent people. In the days preceding Weissberg's arrest, the Ruhemanns helped him considerably, both financially and morally. Yet after his arrest...

From A. Koestler's preface to *The Accused*:

> ... Professor Shubnikov was later on to testify that Alex had tried to recruit him for the Gestapo, which offer he only refused because he (Shubnikov) had allegedly been in the service of a German espionage organization since 1924. Our neighbors and most intimate friends were Martin and Barbara Ruhemanns, of whom the latter, when I asked her to help Alex, affirmed that she had always known him to be a counterrevolutionary saboteur. Every member of that happy band of scientists who used to come in after dinner to play cards or drink tea, stood up after Alex's arrest and denounced him. They were neither cowards nor inferior human beings [but] ...

László Tisza (1907-2009), already mentioned, managed to get out alive, and later became a prominent American physicist. Much has been written about him in connection with Edward Teller, with whom he was on friendly terms since the end of the 1920s. Tisza graduated from Budapest University, then studied in Göttingen. In 1932 he was arrested by the Hungarian government as a communist sympathizer and spent 14 months in prison. After his release, he moved to Kharkov, and worked in the theory group of Lev Landau up to its demise in 1937. Tisza managed to escape from Kharkov to Paris. In 1938, he proposed a two-fluid model of helium-II, explaining the occurrence of superfluidity. He emigrated to the US in 1941 and was a distinguished Massachusetts Institute of Technology professor until 1973. Luck was on his side, you might say ...

From the transcript of a 1987 interview with Tisza:

In Copenhagen Teller met Landau, and told Landau about me, and at that time Landau was in Kharkov in the Ukrainian Physical Technical Institute, which was at the beginning a very promising Soviet institution. Landau was a very interesting inspiration. Teller said,

"Would you take him?"

"Why not?"

Then I got the invitation in June, 1934, to visit Kharkov. There was an international meeting, with a number of British, and French physicists participating. I don't think any Americans went there, but Bohr himself went, and Solomon, who was a son-in-law of Langevin...

Landau offered me a research fellowship to be an *aspirant*, you know, the term *aspirant*, it's a doctoral candidate. Although I had my PhD, but Russia had a completely different standard. Doctor was a very high degree. The candidate degree corresponded to a PhD. In January, 1935, I arrived in Russia. In 1934, when the whole deal was struck, the atmosphere was relatively relaxed. It was after the big collectivization crisis, but in 1934 there was a good harvest, the first good harvest in many years, and the general feeling was that things are going better... By the time I came there in January, the situation was not nearly as relaxed...

About the same time Landau wrote several papers on an innovative theory of the phase transitions that Ehrenfest had recently classified as a transition of second order. By that time I already had a reading knowledge of Russian and I was asked to translate Landau's Russian manuscript into German for publication in the *Physikalische Zeitschrift der Sowjetunion*, a journal which was edited at the UPTI...

Fritz Houtermans was there. Originally a theorist in astrophysics, he worked there in nuclear physics. He started off with neutron physics. And Fritz Lange, a German from a German electric company. He had a famous work on ... aiming the lightning in the Alps and devising high tension. He developed a method of taking condensers in parallel,

> charging them and switching ... to get high voltage. That was a method of getting particle accelerators, and he built one in Kharkov...

Friedrich (Fritz) Lange was one of the authors of the 1940 memo "The Use of Uranium as an Explosive," which was drafted and sent to the Soviet government, as previously mentioned. Born in Berlin in 1899, he worked on his thesis under the supervision of Walther Nernst at the University of Berlin, where he later received a research position. He was an active member of the German Communist Party. In 1935, fleeing from the Nazis and invited by UPTI's director, he moved to Kharkov, where he was appointed head of the laboratory of high-voltage studies. The 1937 disaster at UPTI did not touch him. I do not know why. In the summer of 1942, before the German occupation of Kharkov, Fritz was evacuated to Ufa together with the Institute. There he developed a method for the separation of uranium isotopes using centrifuges, as is still done today (Uranium-235, suitable for the A-bomb, represents a small fraction of the material in raw uranium ore, which consists mostly of Uranium-238). In 1952, Lange was allowed to return to Dnepropetrovsk Polytechnic Institute, and in 1954 was appointed head of a department of the Electrical Engineering Institute in Moscow. Lange failed in his first attempt to get permission to leave for East Germany (the now defunct GDR) in 1957, but the second attempt was a success. On March 19, 1959, Fritz Lange left the USSR for good. Lange was the only foreign scientist working in the Soviet Union under Stalin who was not subject to any reprisals. He died in 1987, two years before the fall of the Berlin Wall.

Konrad Weisselberg. Born in 1905 in East Romania into a Jewish family. Higher education, Doctor of Chemistry, a member of the German Communist Party. Arrived in Kharkov in 1934 at the invitation of a coal industry research center. In 1936 he was hired by UPTI and became Weissberg's neighbor and a good friend. In Weissberg's *The Accused* Konrad Weisselberg appears as Marcel. He was married to a Ukrainian girl, Anna Mykalo (Lena in Weissberg's

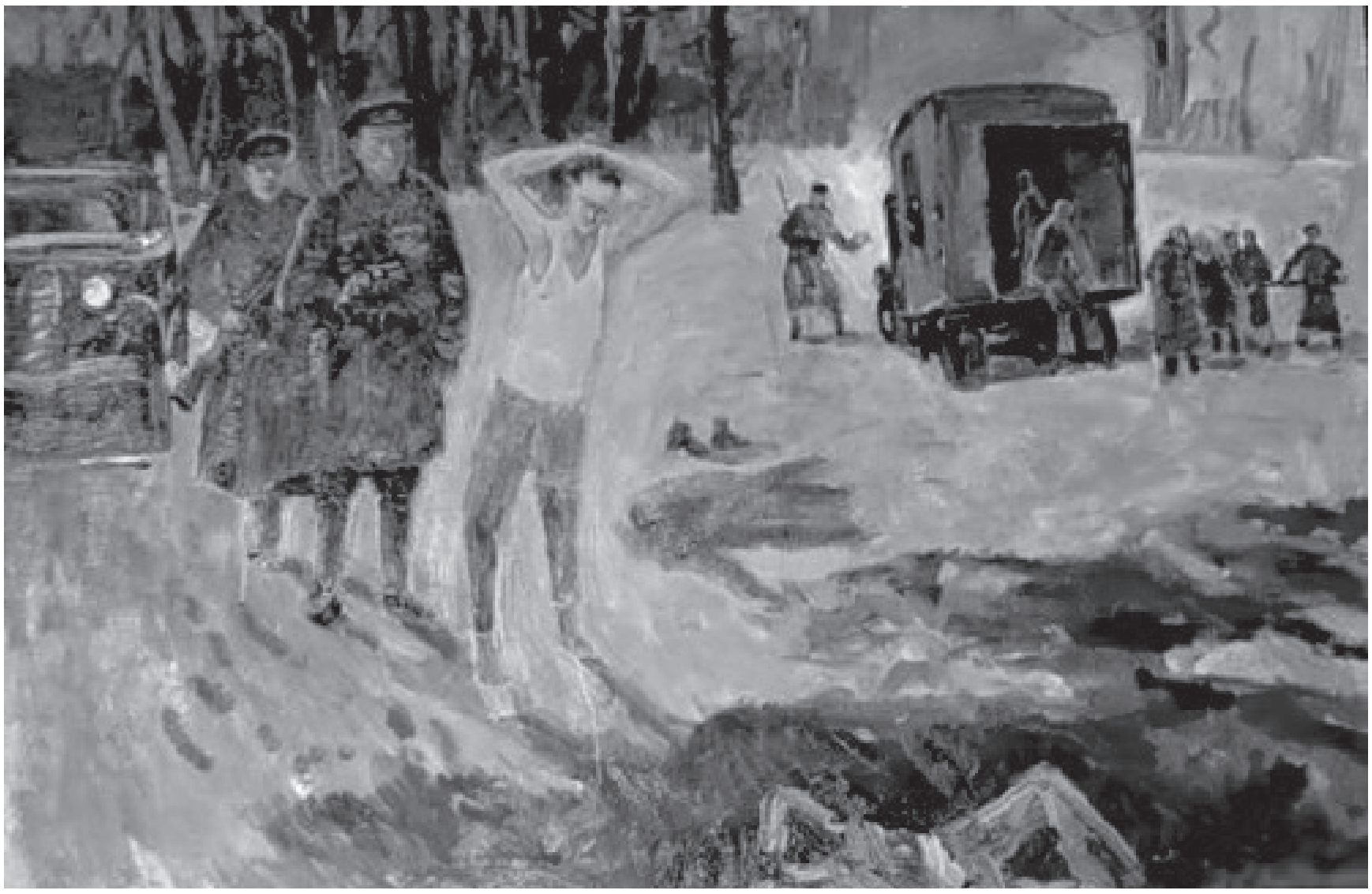

**Egil Veidemanis, Butovo NKVD Execution Range (near Moscow). In 1937-1938 over 20,000 ``enemies of the people” were shot there by firing squads.**

*The Accused*), but this did not help him.

In February 1937 Konrad Weisselberg denounced his Austrian citizenship; he was arrested on March 4, 1937. Weisselberg was indicted on charges of “establishing a criminal contact with the German intelligence agent A. Weissberg and becoming a member of his counter-revolutionary group.” On October 28, 1937, an NKVD *troika*,[14] sentenced him to death; he was executed in December of the same year [9].

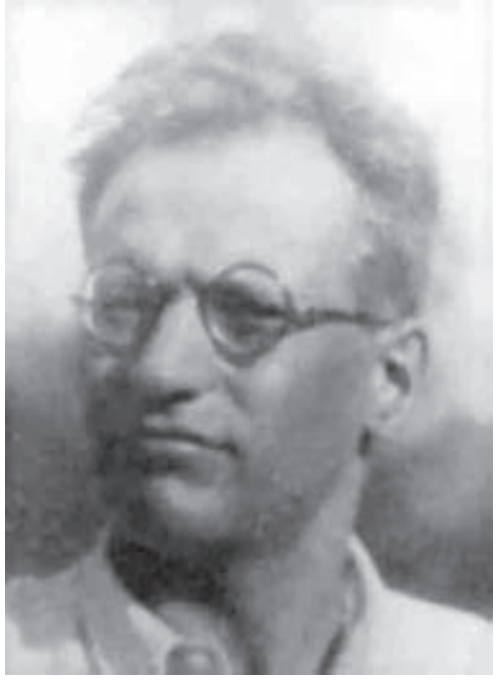

**Konrad Weisselberg in Kharkov**

We can see that the Houtermans’ fates were not the direst...

[14] NKVD *troikas* in the Soviet Union at the time of the Great Purge were institutional commissions of three persons who issued sentences to people after simplified, speedy investigations and without a full trial. These commissions were in fact a subdivision of NKVD used as an instrument of extrajudicial punishment.

— 11 —

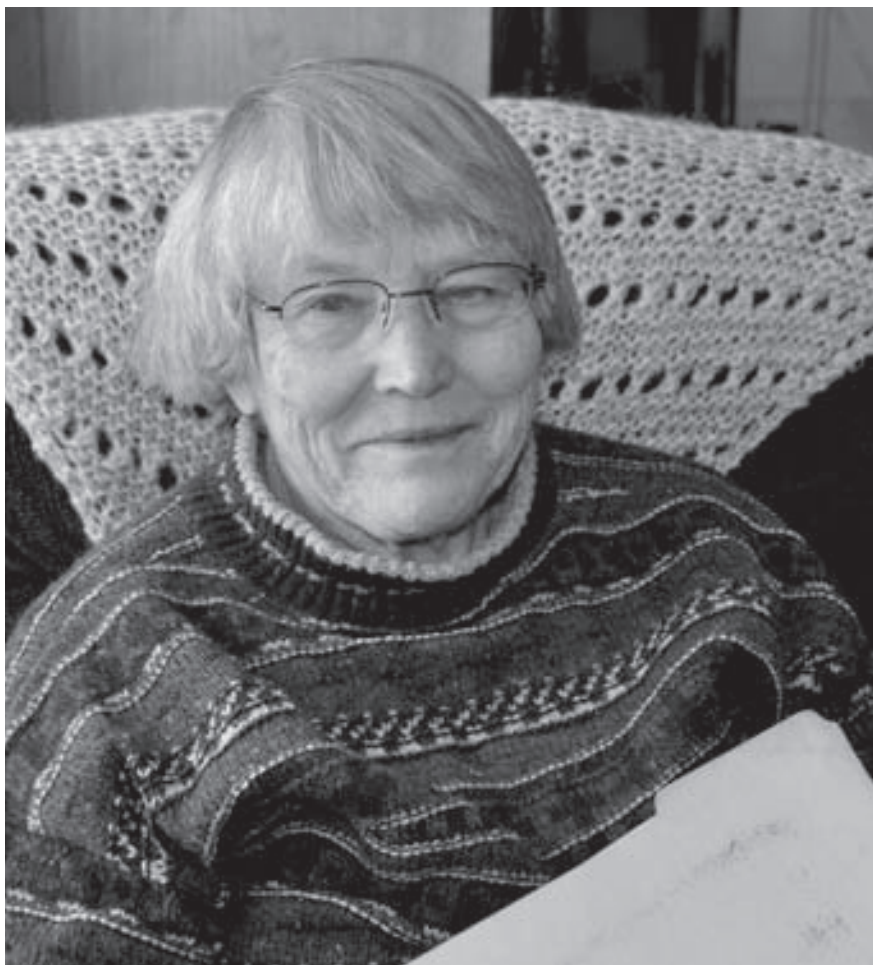

Giovanna Fjelstad, 2014

In search of Charlotte Houtermans' diaries, on one sunny Sunday in March of 2014 I hit the road southbound, toward Northfield, Minnesota. Charlotte was born in Bielefeld, Germany, on May 24, 1899. In the late 1980s she settled in the house of her daughter Giovanna Fjelstad in Northfield, where she died on January 6, 1993. What a long and eventful life...

Northfield, a tiny town (twenty thousand inhabitants) in Minnesota's heartland, is home to the College of St. Olaf and Carleton College, two reputable private colleges. For many years Giovanna Fjelstad, now retired, was a professor of mathematics at the College of St. Olaf.

Giovanna was born in Berlin, Germany, on April 13, 1932. Thus when brought to Kharkov she was three years old. She still remembers a little bit of Russian. In Giovanna's hospitable home we heard some stories from her past, and she found two boxes filled with yellowed pages handwritten or typed by Charlotte Houtermans long ago.

Charlotte's diaries and notes were in some disarray, with some pages unnumbered and undated. At the top of the box I saw a story entitled "Marussya" — a few pages clipped together — which was apparently prepared for publication but seemingly has never been published. It was virtually impossible to find there the parts of the diary that Charlotte had given to Victor Frenkel in the early 1990s during their encounter. Giovanna's daughter, Annika Fjelstad, generously offered her help. In a few weeks I received in the mail a

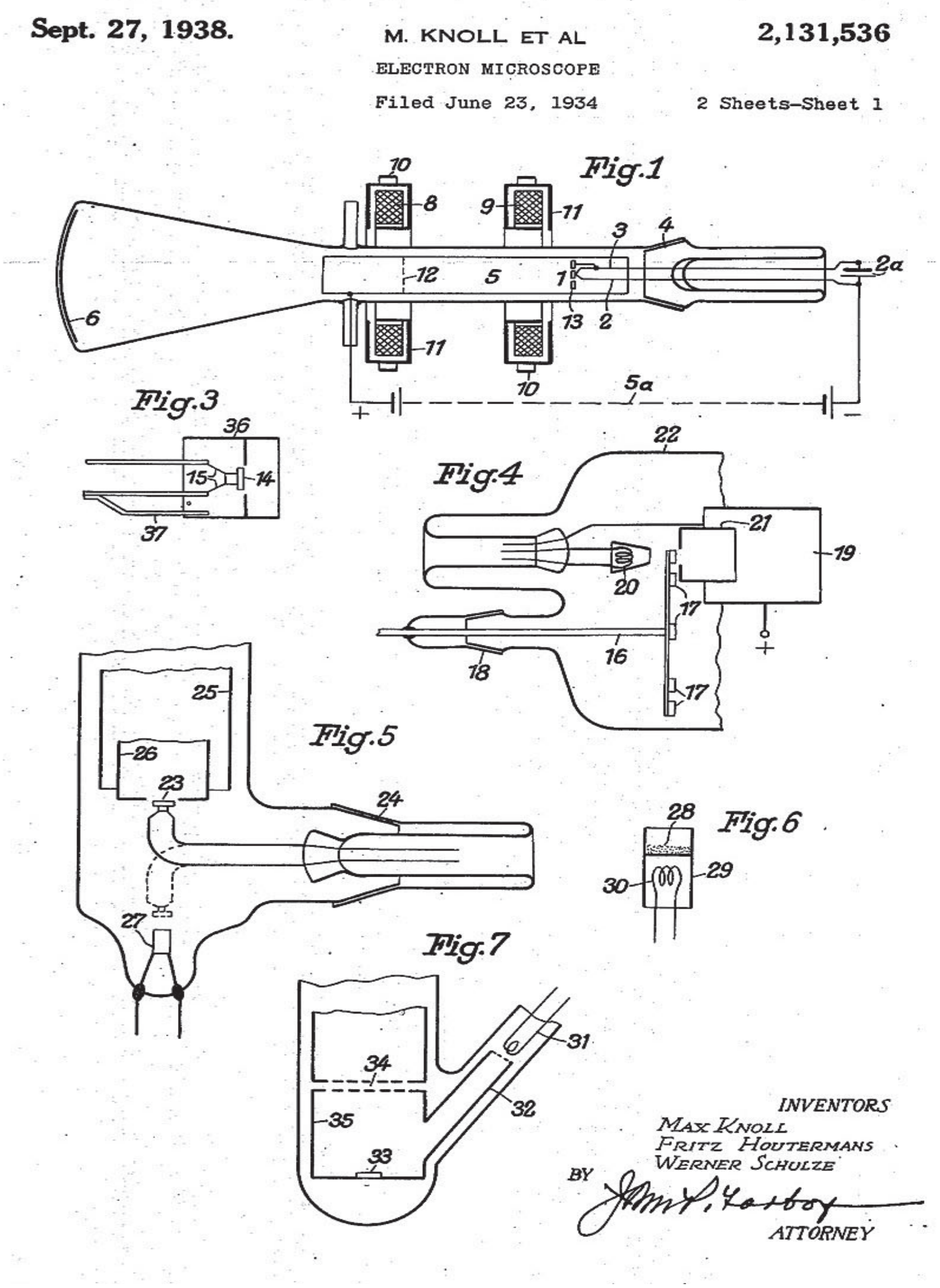

**Knoll-Houtermans-Schulze patent claim (1934) Electron microscope invention.**

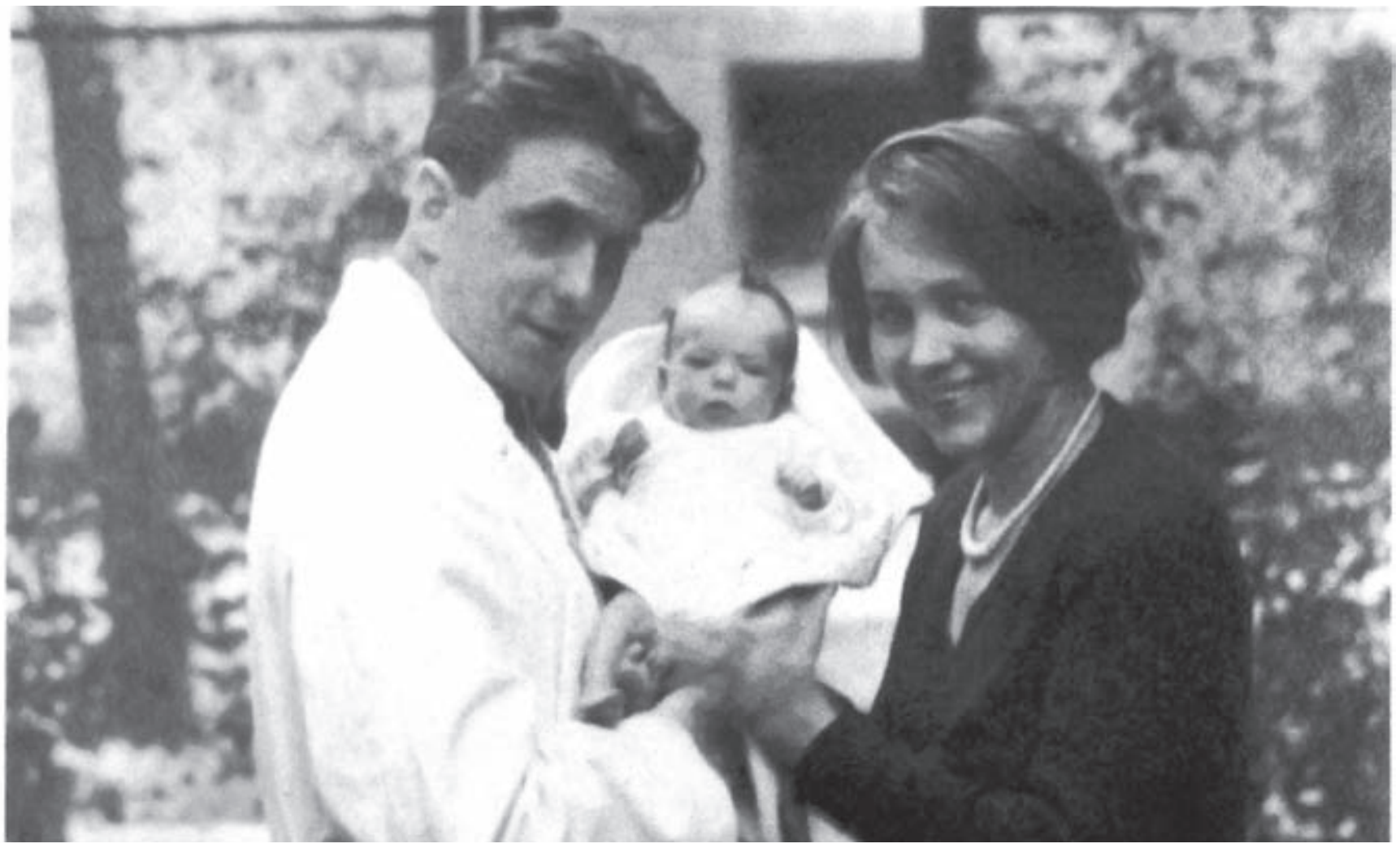

**Fritz, Charlotte and Giovanna in Berlin, 1932**

typewritten manuscript tediously assembled by Annika from small pieces. Thank you, Annika!

— 12 —

Again and again, the Houtermans family had to call new places home.

1933. Fritz received a telegram from his mother, Elsa Houtermans, in which she wrote that she had made up her mind to emigrate to the United States[15] and asked her son to urgently come to see her in Vienna with Charlotte and their daughter.

Fritz and Charlotte stayed at his mother's house, the Star of David and the word "Juden" scrawled in yellow paint on its windows. Elsa's home after the pogrom was in shambles: broken busts of Heine, Goethe, Beethoven, paintings trampled, cabinets open, family documents scattered on the floor [13].

Fritz and Charlotte helped Elsa to pack her suitcases and trunks.

[15] Elsa Houtermans made it to the United States only in 1935, based on an invitation from one of her pupils [2].

Little Giovanna sat quietly with her toys, huddled in a corner of the couch.

"My children," said Elsa, as her descendants tell it, "I would not want you to remain in Germany. Dark ages are approaching in Germany. Fritz, I do not want you and our girls to risk. You have to leave before it's too late."

"Where?"

"This, my children, is up to you," replied Elsa, "but you need to decide quickly. I know what I'm saying ... I've sold my house and I'm leaving you some money. Try again in England, in Russia, but best of all, come to me in America."

1939. Charlotte and her two children, Giovanna and Jan, had to leave England, where her financial situation was dire — there was no way for foreigners to earn a living in England. Thank God, her mother-in-law, now based in the US, helped her to obtain an American visa. Charlotte managed to resume her employment with Vassar College in Poughkeepsie, New York. She was lucky to obtain a research scholarship there through old friends, Edna

-i-

Brione: July 1966.

Again I am looking down onto the lake. I found a bench on the road above the village, half hidden under a tree so that I can write in the pleasant half-shade. Today the water is slate gray and brownish and the mountains have sharp silhouettes. It is peaceful and beautiful and I am at rest.

Many things I deem to have left behind me beyond the mountains. It may be good in this slightly detached mood, in which I find myself, remote from any interference, to take stock and to draw the final line under my life with F., if I can. Somehow it is doubtful, that this should be possible. Too much was involved, too many invisible Threads still tie me to him in ways which I really do not quite grasp. Our life might have been easier if I had understood him. May be I never even knew him, never really found out a single thing about him. I believed in him, I loved him, I invested much love and thought to help him achieving what I thought at the time he wanted to accomplish and what I firmly believed was a very worthwhile goal.

The last weeks in Berne were very confusing, not the fact that Bamsi was packing up her house in Muristrasse I8 and still entertained Pit and

**The front page of the last chapter of Charlotte's diary**

Carter and Monica Heaba. Only nine years before, she had left this college bound for Europe, madly in love, with high hopes for a future happy life in Germany... In 1946 she was granted American citizenship. Approximately at the same time Maria Goeppert-Mayer recommended Charlotte Houtermans for the position at the Sarah Lawrence College.

Fritz divorced Charlotte in 1943. He had a complicated personality. As I learned from various sources — and Giovanna confirmed this — Fritz Houtermans was married four times. Charlotte was his first and third wife in four marriages. Their two children were daughter Giovanna (born in Berlin, 1932) and son Jan (born in Kharkov, 1935). A German law enacted in the Hitler era allowed simplified divorces in the absence of a spouse due to wartime separation. Fritz made use of this law in 1943.  In February 1944, Fritz Houtermans married Ilse Bartz, a chemical engineer with whom he worked during the war and even published a paper. Fritz and Ilse had three children, Pieter, Elsa, and Cornelia. In August 1953, again with Pauli standing as a witness, Charlotte and Fritz were married anew, only to divorce again after a few months. In 1955, Fritz Houtermans married Lore Müller, a sister of his stepbrother, Hans. Lore and Fritz had a son, Hendrik, born in 1956.

Giovanna met her father for the first time after Kharkov only in 1950, when she was 18.

— 13 —

Fritz Houtermans came to the USSR for the first time in 1930, to attend the All-Union Congress of physicists in Odessa. In fact, it was a rebranded Seventh Congress of Russian physicists, organized by the Russian Association of Physicists for August 19-24, 1930, in Odessa. In attendance were over 800 delegates, with two hundred talks covering all branches of physics. Among the foreign participants, except Houtermans, were Sommerfeld, Pauli, F. Simon, R. Peierls. Charlotte Riefenstahl was there as well.

For the city of Odessa this Congress was a great event. Plena-ry sessions were held in the building of the City Council, and the opening was broadcast on the radio. The city authorities took good care of the participants of the Congress, providing them with the best hotels. Delegates could travel free on trams. All sorts of entertainment were organized: tickets to theaters, cinemas, excursions, etc. However, the most popular entertainment in the free hours between the morning and evening sessions was the famous Odessa beach. R. Peierls, who came to the Soviet Union in June 1985, showed several 55-year-old pictures during his talk at the Leningrad Physico-Technical Institute. Judging from the expressions on their faces, Pauli, Frenkel, Tamm and Simon, captured in bathing suits, continued scientific debate even on the beach [14].

The Congress organizers arranged a boat trip on the ship "Georgia" to Batumi, Soviet Georgia, for the participants. Apparently, during this trip the ship made a stop in Sukhumi, where the wedding of Charlotte and Fritz occurred. In my conversation with Giovanna Fjelstad, she said she had no memory of her mother Charlotte ever mentioning Sukhumi, but that Batumi had been mentioned many times.

— 14 —

I do not know who invented the nickname Fisl. But that's how family and friends would address Fritz Houtermans. They say that "Fisl" is a play on the German word "Wiesel," distorted by an Austrian accent. Wiesel means weasel, and in general, a nimble, fun animal, which — they say — fit Houtermans' character. Fritz-Fisl himself was good at inventing funny nicknames. Thus Giovanna became Bamsi and family friend Charlotte Schlesinger Bimbus...

Fritz had a sharp sense of humor. One of his colleagues, Haro von Buttlar, collected anecdotal stories told by Houtermans and published them in a 40-page book [15].

There is a legend that it was Fritz Houtermans who applied the name "Martians" to seven of the twentieth century's most outstanding Hungarian scientists, Theodore von Kármán, George de Hevesy,

Michael Polanyi, Leó Szilárd, Eugene Wigner, John von Neumann, and Edward Teller, because nobody in the West could understand their language. One can find this attribution in the "pages" of Wikipedia. Unfortunately, this legend apparently has no ground. In fact, it was a fellow Martian, Leó Szilárd, who jokingly suggested that Hungary was a front for aliens from Mars [16].

— 15 —

In editing the English translation of the Frenkel's book I had to make some decisions. Comparing the Russian edition with Charlotte's diaries, I noted that in a few instances V. Frenkel abbreviated the original by discarding words, sentences and even some paragraphs. On the other hand, in one or two instances Frenkel's Russian translation contains details which I could not find in those pages of Charlotte's diaries that I had in my possession.

For this English edition, I decided to present Charlotte's diaries the way they had been written. Where necessary I added footnotes for explanations. In general, Frenkel's book is a masterpiece that succeeds as both a scientific biography and a captivating read. However, many background events that may be sufficiently well-known to the Russian reader but may remain relatively obscure to the Western reader needed accounting for. My preface as well as numerous footnotes scattered throughout the text hopefully fill the gap and introduce intriguing layers of additional research.

Moreover, I decided to supplement Frenkel's book by adding three appendices. The first one is a short excerpt from E. Amaldi [2] which covers the last years of Houtermans' life. The second appendix is a chapter, "Odessa 1930," from the book *Fisl, or the Man Who Overcame Himself* by B. Diakov et al. [13], describing F. Houtermans' first visit to the Soviet Union. The latter book is based on materials in Frenkel's *The Last Works* [17], published posthumously.

The third appendix is a slightly abridged piece of personal writing by Charlotte Houtermans, containing a wealth of hitherto unpublished material. With some hesitation I decided to omit

Diakov's brief preface to the Russian edition, as this introductory statement does not directly relate to the body of the book. I also omitted a foreword by Academician Zhores Alferov.

A list of F. Houtermans' scientific publications can be found in the above-mentioned book by E. Amaldi or in the Russian version of Frenkel's book. It is also reproduced in the German version [18] of Frenkel's book.

— 16 —

Part II is devoted to Yuri Abramovich Golfand (1922-1994), whom I knew personally. We were not close friends, because of the age difference, but I always felt an almost irrational attraction to him. Golfand was a frequent participant in the ITEP[16] theory seminars. I used to bump into him in the corridors of ITEP regularly. At first I did not know who this small man was, with his warm eyes and kind smile. So I asked my thesis adviser, Prof. B.L. Ioffe. Ioffe lowered his voice to a whisper and replied that this was Golfand, the discoverer of supersymmetry. Later, whenever he spoke of him, Ioffe would automatically lower his voice even if we were alone in Ioffe's office. This would wordlessly emphasize that Golfand, as a *refusenik*, was a nonperson.

Everybody who knew Golfand remembers his smile and his eyes. Usually he looked a little bit out of touch with reality, decoupled from the surrounding world, with thoughts directed within rather than without.

After Yuri Abramovich died in 1994, flashes of memory often revived his smiling face in my mind. I could not forget it. During his lifetime he had no opportunity to travel to the West, and therefore his early works — representing the inception of supersymmetric field theory — were known only to a few experts, pioneers in this subject, and his (and Evgeny Likhtman's) role in this inception was underappreciated. In 1999 I edited the anthology *The Many Faces of*

[16] The Institute of Theoretical and Experimental Physics in Moscow.

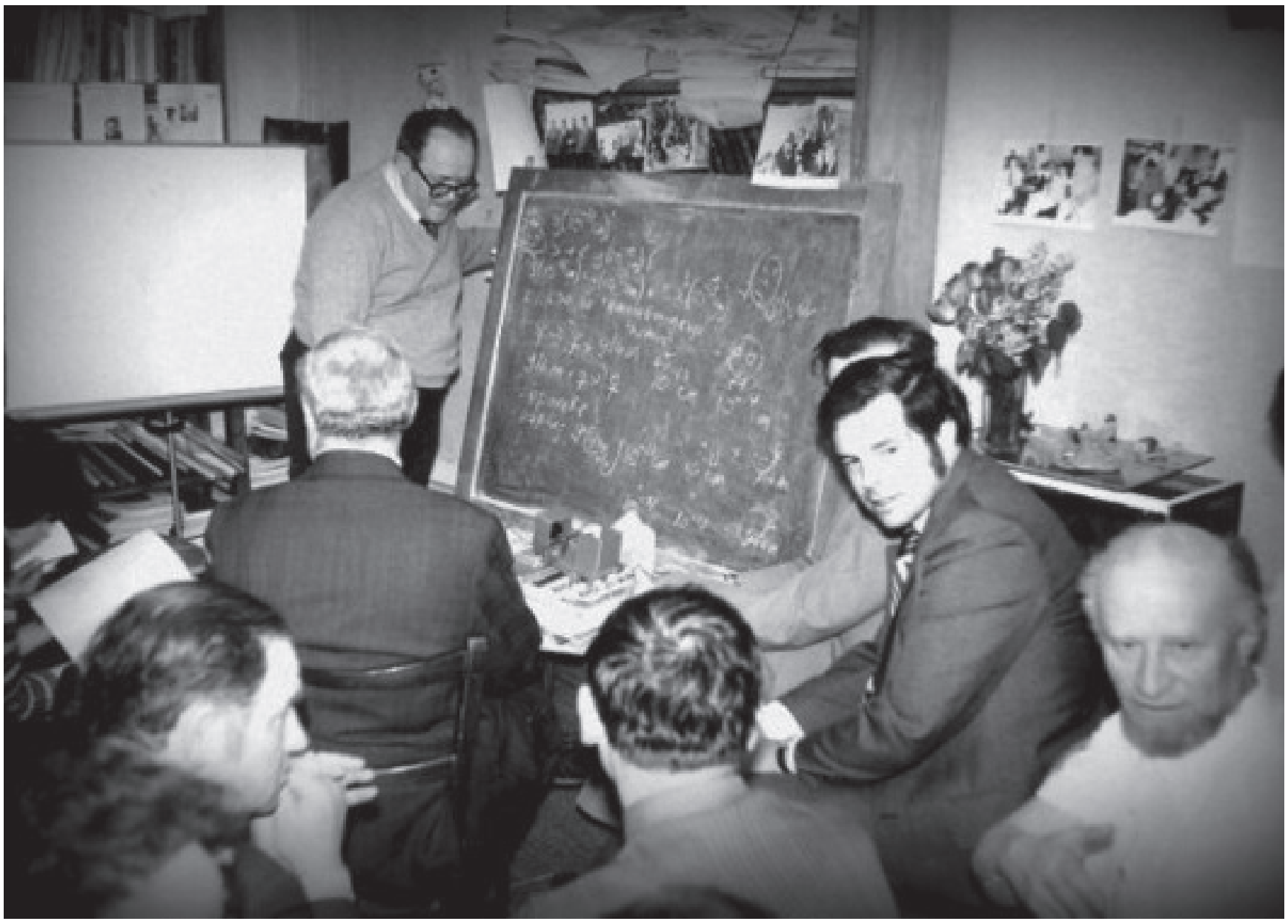

**Physics seminar of Moscow *refusniks* in a private apartment. In the center-right, looking in the camera, is Kenneth Wilson, a guest from the US and the future Novel Prize winner.**

*the Superworld* [19], dedicated to the memory of Yu. Golfand.

My intention was to correct an historical injustice. Boris Bolotovsky in his essay mentions that it was this volume's appearance which led him to write his "Memories of Golfand" (see Part II of this book). In the same year, I (with G. Kane) published the book *The Supersymmetric World* [20], which presented recollections from some of the founding fathers of supersymmetry, as well as from Golfand's widow, Natasha Koretz.

On October 13-27, 2000, the William I. Fine Theoretical Physics Institute (University of Minnesota) hosted the symposium and workshop "Thirty Years of Supersymmetry." During this event, some of the pioneers who opened the gates to the superworld in the early 1970s met face-to-face for the first time ever. Among the guests of the symposium were Natasha Koretz, Evgeny Likhtman, Vladimir Akulov, Vyacheslav Soroka, Pierre Ramond, Jean-Loup Gervais, Bunji Sakita, Pierre Fayet, John Iliopoulos, Lochlain тониане взаимодействия. Однако, при произвольной локальной плотности гамильтониана $H^I(t)$ плотность оператора $W^I(t)$ не будет, вообще говоря, локальной за счет интегрирования в (6) по времени. Поэтому, чтобы получить физические следствия, мы обобщим понятия инвариантности теории относительно алгебры (или группы) на случай, когда операторы алгебры сами могут существенным образом зависеть от взаимодействия. Мы будем требовать, чтобы пространственная плотность операторов алгебры была локальной функцией от операторов поля [1]. В результате этого требования подинтегральные выражения в (6) должны быть полными производными по времени, и соотношения (6) превращаются в уравнения для определения операторов $H^I(t)$ и $W^I(t)$.

Уравнения (6) сводятся к системе линейных однородных уравнений для постоянных коэффициентов, которые вводятся в качестве неопределенных констант связи в наиболее общий вид гамильтониана взаимодействия. Эту систему уравнений удалось решить в случае, когда $H_1(t)$ является произведением трех полей, два из которых преобразуются по представлению (2) и комплексно-сопряженному (2), а третье – по представлению (3). Система уравнений (6) в этом случае имеет единственное решение, а отличны от нуля лишь операторы $W_1(t)$, $H_1(t)$, $H_2(t)$. Зная точный вид гамильтониана в представлении взаимодействия, можно восстановить по нему лагранжиан в гейзенберговском представлении:

$$L(x) = (\partial_\alpha \phi^* - igA_\alpha \phi^*)(\partial_\alpha \phi + igA_\alpha \phi) - m^2\phi^*\phi + (\partial_\alpha \omega^* - igA_\alpha \omega^*) \times$$
$$\times (\partial_\alpha \omega + igA_\alpha \omega) - m^2\omega^*\omega + \frac{i}{2}\bar{\psi}_1 \gamma_\alpha \overleftrightarrow{\partial}_\alpha \psi_1 - m\bar{\psi}_1\psi_1 - g\bar{\psi}_1\gamma_\alpha\psi_1 A_\alpha +$$
$$+ \frac{i}{2}\bar{\psi}_2\gamma_\alpha\overleftrightarrow{\partial}_\alpha\psi_2 - \mu\bar{\psi}_2\psi_2 - \frac{1}{2}(\partial_\beta A_\alpha)^2 + \frac{\mu^2}{2}A_\alpha A_\alpha + \frac{1}{2}(\partial_\alpha X)^2 - \frac{\mu^2}{2}X^2 +$$
$$+ g\mu(\phi^*\phi - \omega\omega^*)X - \frac{g^2}{2}(\phi^*\phi - \omega^*\omega)^2 + \sqrt{2}g(\bar{\psi}_1\bar{s}\psi_2\phi + \bar{\psi}_2\overset{+}{s}\psi_1\phi^*) -$$
$$- \sqrt{2}g(\psi_1^c\bar{s}\psi_2\omega^* + \bar{\psi}_2\bar{s}\psi_1^c\omega) . \qquad (7)$$

Таким образом, получена модель взаимодействия квантованных полей с несохранением четности, инвариантная относительно алгебры (1).

Физический институт
им. П.Н.Лебедева
Академии наук СССР

## Литература

---

[1] Обоснованию этого постулата, а также сравнению его с обычной формулировкой требования инвариантности теории относительно группы преобразований будет посвящена отдельная работа.

**The last page of Golfand-Likhtman paper (JETP Letters, v. 13, p. 452, recieved by the Editorial Board on March 10, 1971), with the first ever four-dimensional supersymmetric model presented in Equation (7). I suggested to call this model – massive supersymmetric electrodynamics – as the Golfand-Likhtman model.**

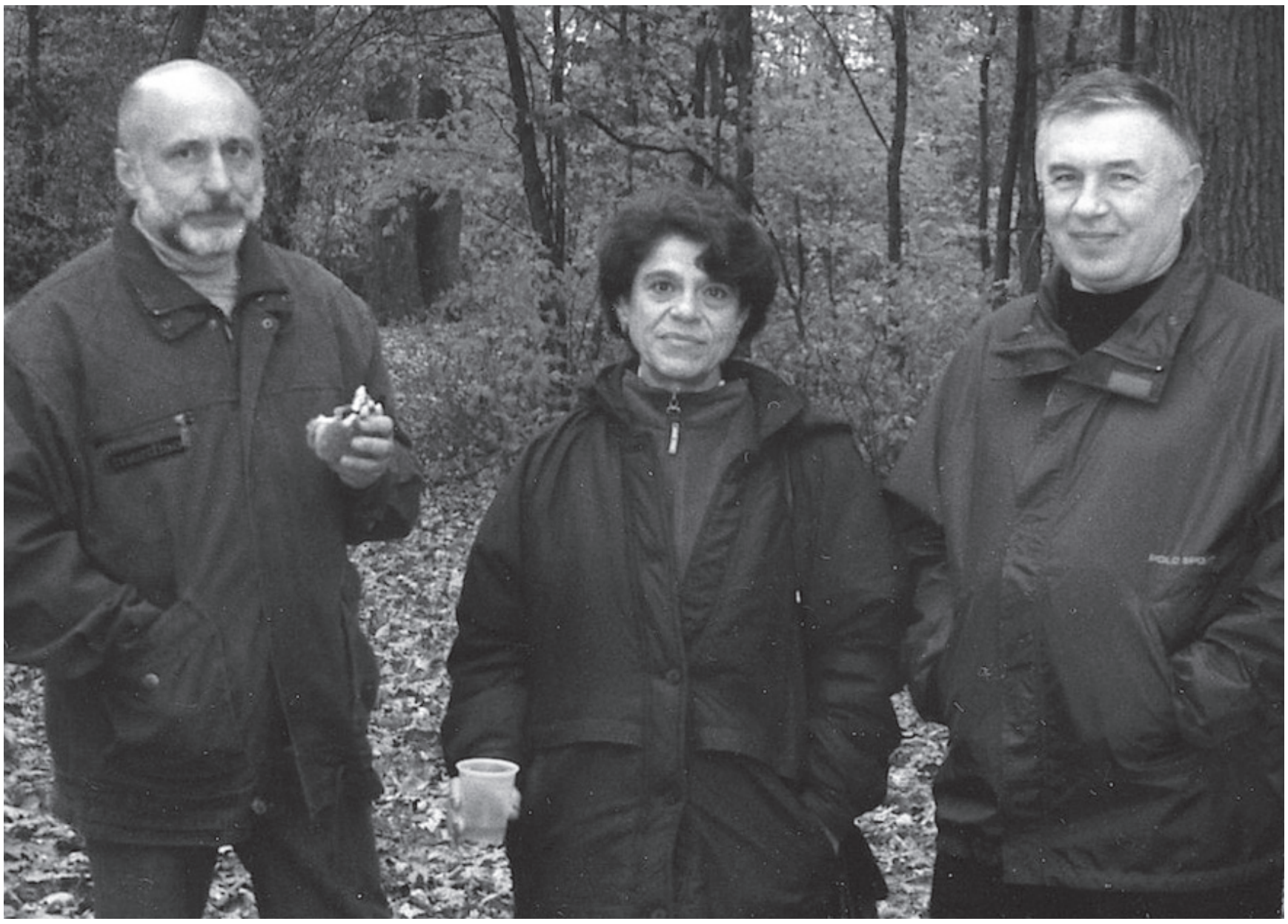

**Evgeny Likhtman, Natasha Koretz, and V. Akulov in Minnesora in October 2000.**

O'Raifeartaigh, Sergio Ferrara, Martin Sohnius, John Schwarz, and others. Later the proceedings of the symposium were published [21]. Some rarely mentioned aspects of the scientific history of supersymmetry are summarized in the introductory part of *Concise Encyclopedia of Supersymmetry* [22].

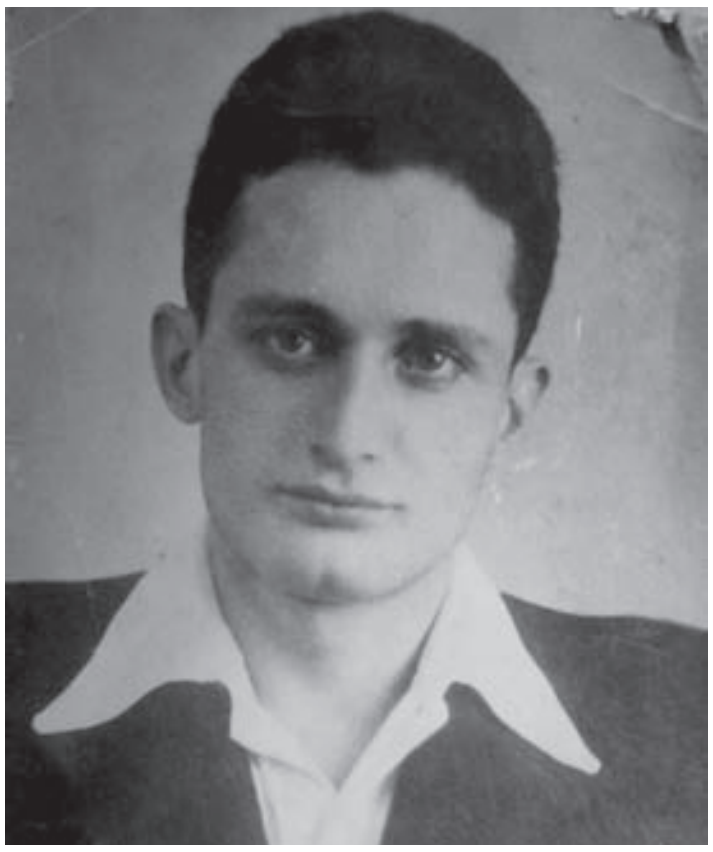

**Yuri Golfand, circa 1940**

Several years ago I received from a Moscow friend a draft of Boris Bolotovsky's essay, which is published in Part II. This is the most detailed account written by a witness of the story of Golfand's expulsion from the Lebedev Physical Institute in Moscow, and his subsequent persecution.

I got in touch with Boris Mikhailovich, and after his essay was published in Russian in the online magazine *Sem Iskusstv* (November 2012) he

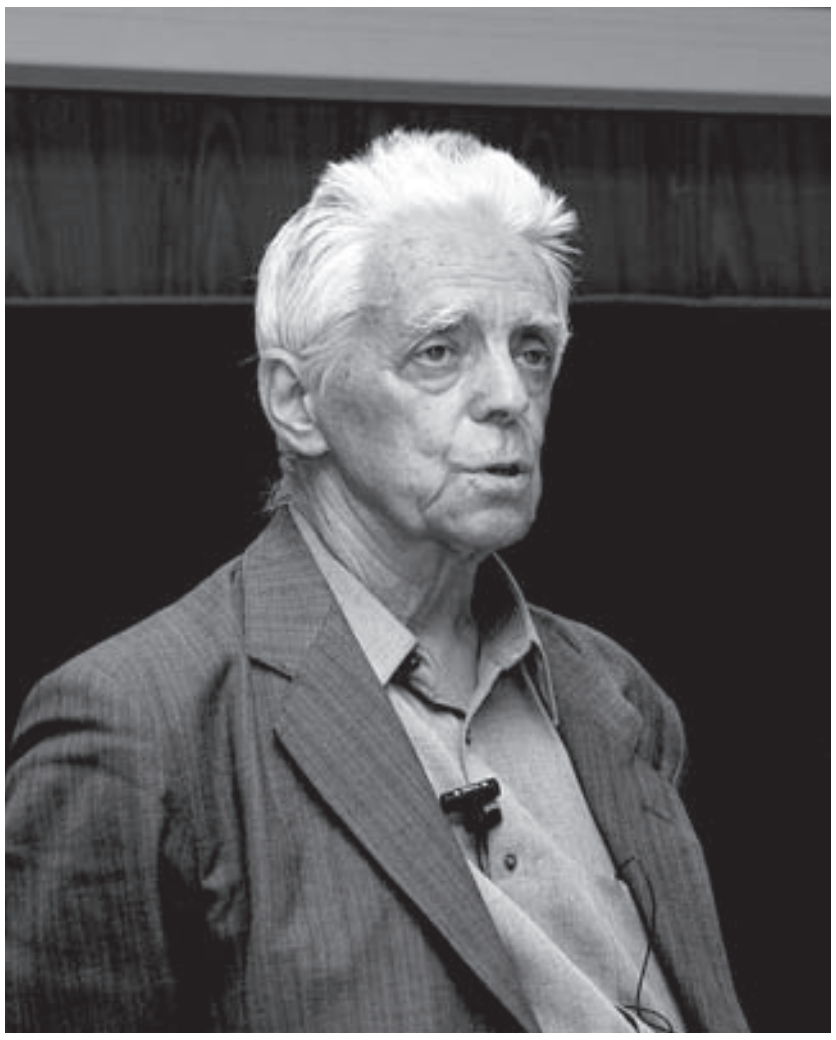

Boris Mikhailovich Bolotovsky

gave me permission for its translation into English. I am grateful to *Sem Iskusstv* editor Evgeny Berkovich for his kind assistance and permission for publication.

In order to clarify a few points in Bolotovsky's essay in the process of editing, I called Natasha Koretz. She kindly informed me that Golfand's biographer in Israel, Boris Eskin, had written a large essay (in Russian) covering Yuri Golfand's life from childhood to his death in 1994. It is the most detailed narrative one can currently find. He kindly agreed to the inclusion of the English version of his essay in this book. Minor abbreviations were made by the Editor. I added a large number of footnotes to make certain details in Eskin's text more understandable to the Western reader.

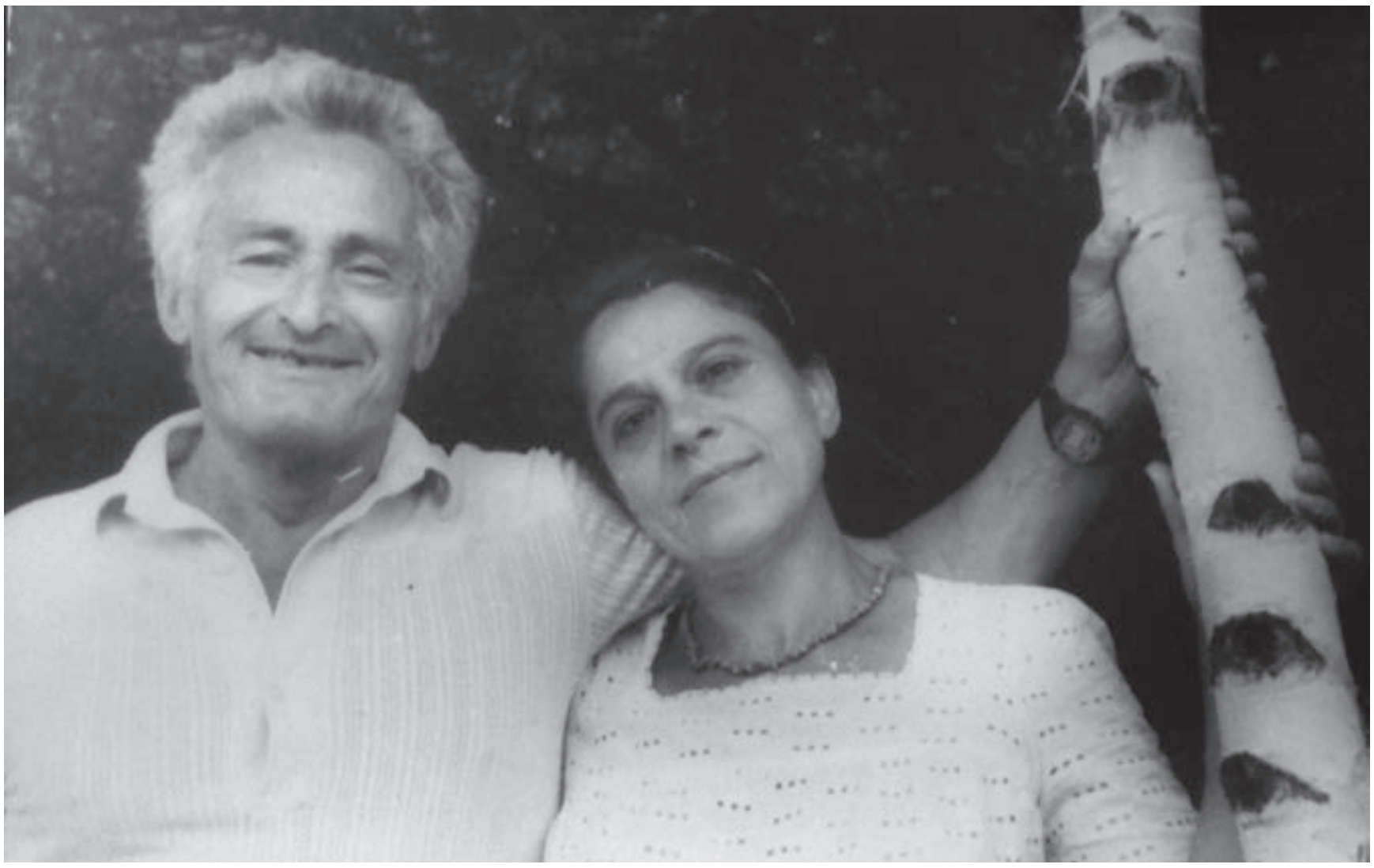

Yuri Golfand and his wife Natasha Koretz, circa 1980.

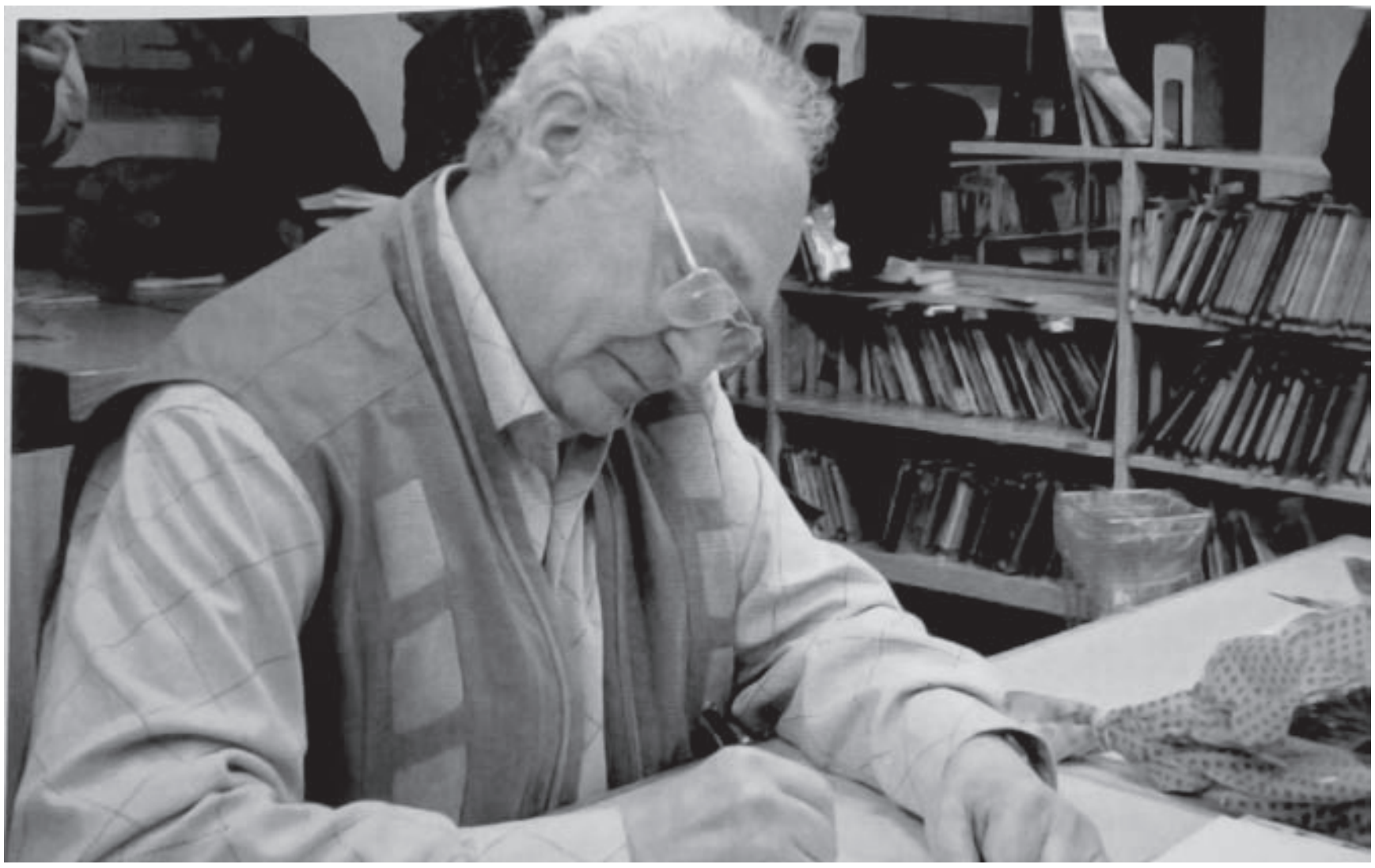

**Boris Eskin**

— 17 —

A digression in Eskin's essay acquaints the reader with the tragic story of Natasha Koretz's father, Moisei Abramovich Koretz (1908-1984). He was born into the family of a Jewish watchmaker, and received his basic education in the Physics and Mechanics Faculty of the Leningrad Polytechnic Institute (1929-1934). There Koretz attended lectures by L. Landau and M. Bronstein.

In 1935, at the invitation of Landau, Moisei Koretz arrived in Kharkov, where he was admitted to the theoretical division of the Ukrainian Physico-Technical Institute and, concurrently, became Landau's assistant at Kharkov University. During this period, Koretz, according to Landau, "proved himself as a capable young physicist" [23]. Soon Koretz became his "close associate and assistant" [24]. Just a half year later, Koretz was fired from UPTI, allegedly "for concealing his social origin," and two weeks later, in the late November of 1935, he was arrested by the NKVD on charges of

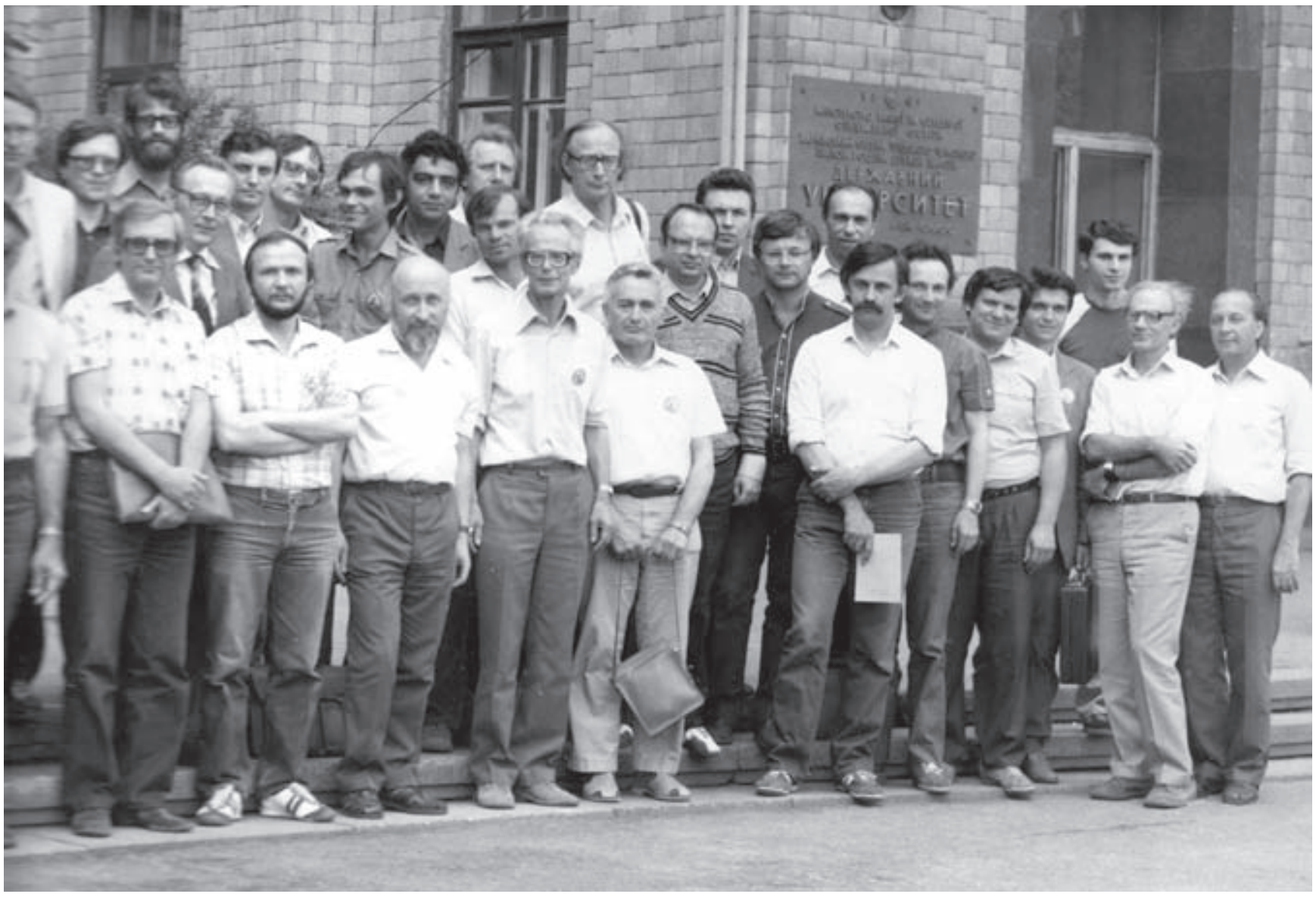

**A conference in Kharkov in 1985. Yuri Golfand is in the center of the front row. To the left of him is Dmitri Volkov, also a founding father of supersymmetry. Behind Golfand is Victor Ogievetsky, one of the pioneers.**

"agitation that led to the failure of defense contracts." He spent eight months in a Kharkov prison anticipating the death penalty. In December 1935 Landau sent a letter in defense of Koretz to the head of the Ukrainian NKVD. Apparently as a result of Landau's and other similar letters from UPTI, the Ukrainian Supreme Court overturned the verdict and Koretz was released.

Natsha Koretz writes [25] about her father: "Moisei Abramovich was not only Landau's student in Leningrad, he also became his most trusted friend, with whom Landau could talk about all the things in the world that interested him: politics, future, love and art... For example, they together invented a classification of female types, and what approach one should apply to each of them for seduction... They also worked out a general classification of human intellect, i.e. people who can create novel things, at most 2%, those who are able to accept these novel things right away 5%, those who are able to discuss problems expressing their own opinions 13%, those who can recognize a problem as such 20%, while the rest — according to this classification — create just `noise.' This and much more bonded them. That's why Dau invited my dad to Kharkov, when he became the head of the Theoretical Department there.

It was no accident that they wrote their notorious leaflet together when they moved to Moscow. They could easily have paid with their lives for it, had it not been for Kapitsa."

Indeed, Koretz, following Landau, moved to Moscow in February 1937. After the NKVD archives were opened in 1991, it became known that in April 1938, Koretz and Landau had written a leaflet stating:

> ...We came to the conclusion that the ... Soviet government is not acting in the interests of the working people, but, rather, in the interests of a narrow ruling group. For the sake of our country it would be beneficial to overthrow the Stalinist regime and to establish a state in the Soviet Union, preserving the collective and state ownership of enterprises, but constructed on the model of the bourgeois-democratic states.

On April 27, 1938, Koretz, Landau and Rumer were arrested by the NKVD. Tortured during interrogations, Koretz entered a guilty plea, and was sentenced to ten years of hard labor for "propaganda calling to overthrow, undermine or weaken Soviet power." He served 14 years in Pechorlag (a branch of the Gulag), near the village of Mezhoga. Why 14? In 1942 he was sentenced to an extra 10 years. Among the charges was an alleged remark made by Koretz in early 1941 about a possible German attack on the Soviet Union. In the indictment, these words were presented as "doubts in the power of the Soviet system." In 1952, after serving 14 years in the labor camp, Koretz was sent into exile until 1958 [25].

In Koretz's NKVD file there is an excerpt from the testimony of the UPTI physicists Shubnikov and Rozenkevich, executed in 1937 in Kharkov on charges of espionage for Nazi Germany. In this testimony they admit to their participation in a counter-revolutionary Trotskyist organization, along with both Landau and Koretz.

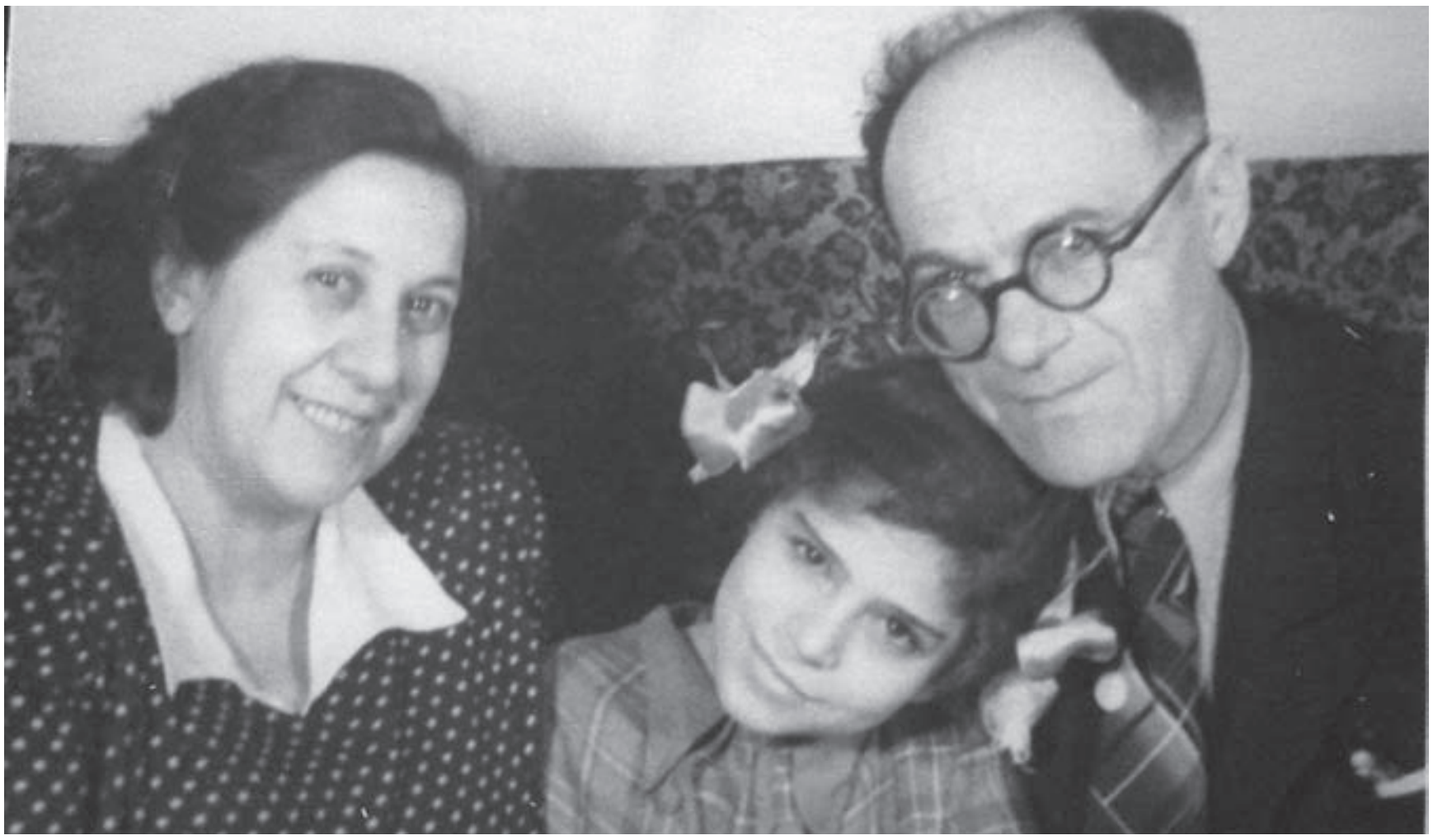

**Natasha Koretz with her parents. Inta, Circa 1954**

— 18 —

The scene of Bolotovsky's essay is Moscow-based FIAN, the Physical Institute of the Academy of Sciences[17] — more exactly, its Theory Department. I should say a few words about this.

FIAN was created in 1934, a few years later than UPTI. Its first director was Academician Sergei Vavilov. The Theory Department was headed by Igor Tamm, a future Nobel Prize winner (1958). He was its Head until his death in 1971, with a five-year break in 1938-1943 when the Theory Department was dissolved. At its inception the FIAN Theory Department consisted of nine members — among them such outstanding physicists as Matvei Bronshtein, Moisei Markov, Yuri Rumer and Vladimir Fock. We will encounter their names in the body of this book. The Great Purge exacted a heavy toll from FIAN's theory pioneers. In 1938 Matvei Bronshtein was sentenced to death, while Yuri Rumer was arrested and sentenced to 10 years as an accomplice of the enemy of the people Lev Lan-

[17] Also known as the Lebedev Physical Institute.

dau. After serving his term he was exiled to Siberia. Tamm himself fell under the NKVD's suspicion.[18] In addition to heading the FIAN Theory Department, he was a professor at Moscow State University. At that time, the Dean of the Faculty of Physics at the Moscow State University was Professor Boris Gessen. Tamm and Gessen knew each other from childhood. When Gessen was arrested, declared an

---

[18]And not for the first time. In the book *Memories of Tamm* [26] Tamm's grandson tells the following story about his grandfather.

In the summer of 1920 Tamm decided to leave Crimea, controlled by the Whites, for Elizavetgrad, which was already controlled by the Reds. Igor Tamm did not take any documents with him, because they weren't suitable for justifying departure from the territory controlled by the Whites and weren't suitable for joining the Reds, either. He successfully crossed the front line... At night, along with a casual companion, he decided to stay in an empty house at an abandoned manor, where they were detained by a Red squad.

Without documents they looked like White scouts and should have been immediately shot. Fortunately, the detachment commander was a dropout student. Tamm said that he had graduated from the Moscow University Department of Physics and Mathematics, but the commander didn't believe him. He smiled darkly and said: "So you are a mathematician! That has to be a lie, right? Let's check it out. Prove Taylor's theorem on representation of a function by its Taylor series. Don't forget the explicit formula for the remainder! If you cope with the task we will set you free. If you fail we will shoot you and your pal right away."

They gave Tamm a pencil, a piece of paper and a candle, dragged a load of fresh hay to the captives and locked them in the room. His companion quickly fell asleep. Igor Tamm could not sleep: there was a guard at the door and time was scarce. He was nervous... Because of this he made a mistake, which prevented him from proving the statement. Nevertheless, he correctly outlined the proof. In the morning, when the Red commander came in, the task was still unfinished, but it was clear that the unsuccessful attempt of proof was written by someone who knew math. Tamm asked the commander to show him the mistake.

"To tell the truth," he answered, "I can't. I dropped out of university three years ago and I have already forgotten everything."

His companion was released, but Tamm remained in captivity. The Whites went on the offensive and the Red squad together with their captive retreated to Kharkov instead of Elizavetgrad. In Kharkov, a soldier was delegated to hand Tamm over to the Cheka. "In the Cheka they will sort things out quickly," said the soldier.

enemy of the people and executed, the NKVD's attention turned to the Tamm family. His brother Leonid Tamm, a prominent chemical engineer, was arrested and sentenced to ten years in the Gulag, where he disappeared without a trace. In an attempt to save Igor Tamm and other theoretical physicists from Tamm's department from inevitable persecution, Vavilov dissolved the Theory Department; it was reassembled only in 1943.

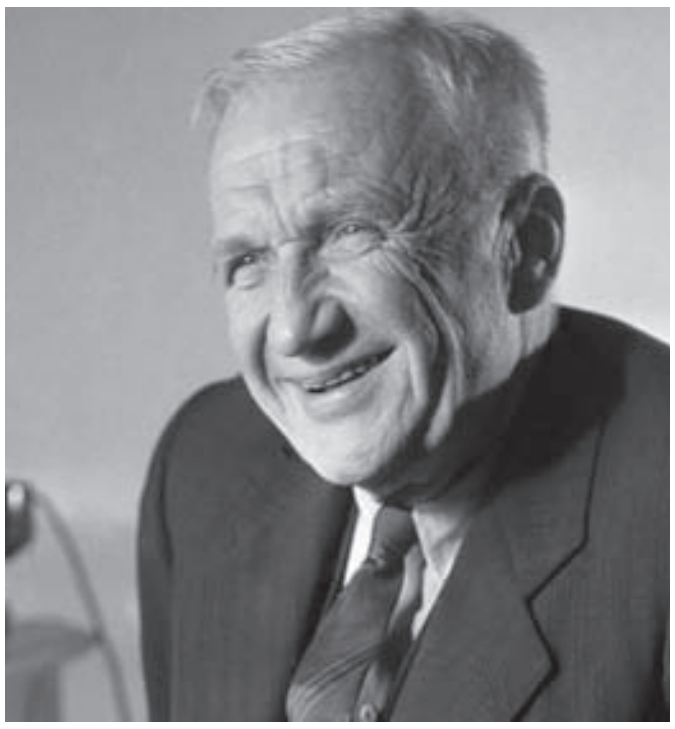

**Academician Igor Tamm**

Vitaly Ginzburg, the future Nobel Prize laureate (2003), became Tamm's student in 1938 and was hired by FIAN in 1940. Both Tamm and Ginzburg participated in the Soviet hydrogen bomb project. But the main player in this project was Andrei Sakharov, another member of the FIAN team.

In the spring of 1945 Tamm acquired a new student, Andrei Sakharov, who at that time was an engineer at the Ulyanovsk ammunition plant. He graduated from the Faculty of Physics, Mos-

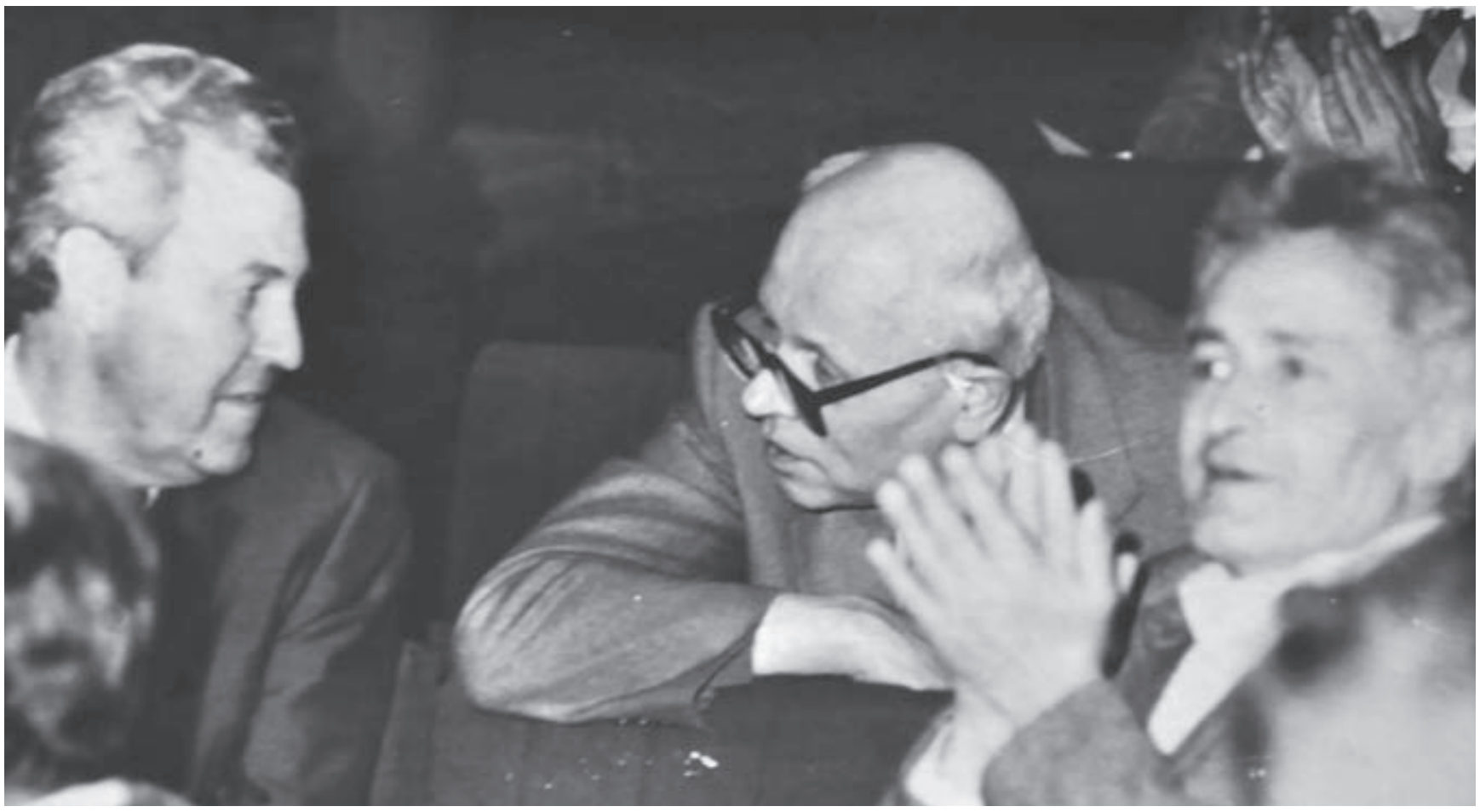

**Andrei Sakharov (center) at a physics conference (1987). To the right is Yuri Golfand, to the left is Alexei Anselm.**

cow State University, in 1942 as one of the best in its class. He was offered a position at the university but declined the offer, wishing instead to join war effort. That's how Sakharov wound up at the munitions plant, where he authored numerous inventions. After the war's end, the desire to engage in fundamental science led Sakharov to Tamm. In 1950 Sakharov was sent to Sarov, the Soviet counterpart of Los Alamos, where he proved absolutely instrumental in the development of the first Soviet hydrogen bomb.

Starting the late 1950s, Sakharov became concerned about the moral and political implications of his work. The turning point in Sakharov's political evolution came in 1967. Shortly after, he authored an essay, *Reflections on Progress, Peaceful Coexistence, and Intellectual Freedom*, which marked the beginning of his dissident career. After this essay was found in public circulation, Andrei Sakharov lost his security clearance and in 1969 returned from Sarov to Moscow and the FIAN Theory Department. Sakharov was arrested on January 22, 1980, following his public protests against the Soviet intervention in Afghanistan in 1979, and was sent, without any judicial procedure, into internal exile in the city of Gorky, now Nizhny Novgorod — a city classified as off-limits to foreigners. Between 1980 and 1986, Sakharov was kept under tight KGB surveillance. Only after the advent of Gorbachev's *perestroika* was he allowed to return to Moscow.

— 19 —

If not stated to the contrary, footnotes in this collection belong to the Editor.

*General Acknowledgments*

First and foremost I am grateful to James Manteith. Not only did he masterfully translate Frenkel's, Eskin's and Bolotovsky's essays in this book from Russian, in many instances he acted as my invaluable adviser both on the style of presentation and the relevance of various excerpts from other sources.

I would like to thank Giovanna, Lars and Annika Fjelstad for providing me with Charlotte's papers, as well as for their inspiring conversations and correspondence, and generous assistance. I am grateful to Anna Lovsky (lovskydesign.com) for the cover design and to Leigh Simmons (leighsimmons.com), who handled other aspects of the graphic design. I acknowledge kind permission from the editors of the book [2], Saverio Braccini, Antonio Ereditato, and Paola Scampoli, to quote an excerpt from Amaldi (see Appendix I).

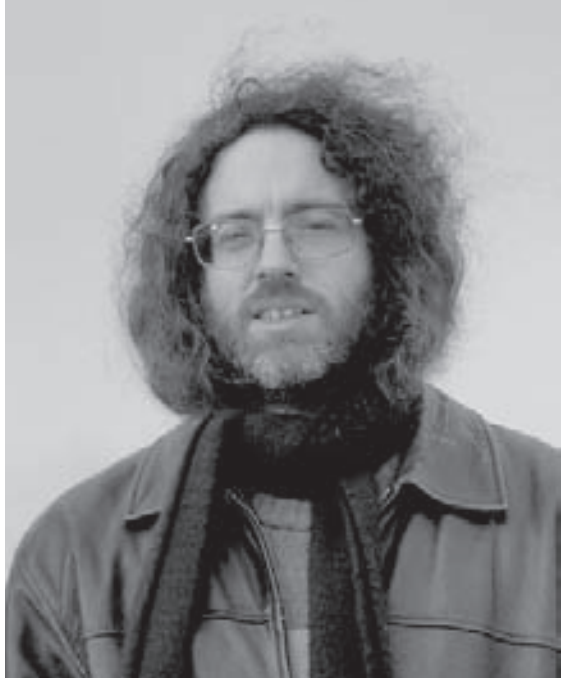

**James Manteith**

I would like to say thank you to Meghan Murray, who was in charge of my grant funds and served as this project's translation business coordinator.

I am grateful to Natasha Koretz for invaluable assistance, advice and encouragement. Helpful correspondence with B. Diakonov, O. Cherneva, E. Nash, H. Leutwyler, N. Tserevitinova, Judith Szapor, Jean Richards, Ben Beynman, Yu. Ranuk, and Lakshmi Narayanan, my friend and World Scientific contact, is gratefully acknowledged. The photographs in this book are taken from family archives (with kind permission of the owners), from my personal archive, and from Frenkel's book. Some pictures were kindly forwarded to me by Professor Dieter Hoffmann from the Max Planck Institute for the History of Science in Berlin, Germany, who published the German translation of the Frenkel's book in 2011 [18], with a number of additional photographs absent from the book's original Russian edition.

May 2015
Minneapolis

*Disclaimer: This book is* not *an academic monograph on the history of science. The Editor made only a limited attempt at checking facts presented by the writers of the essays included in this collection. The opinions expressed by the writers do not necessarily coincide with the opinion of the Editor.*